%% file: main.tex
\documentclass[manuscript,screen]{acmart}

\usepackage{colortbl}
\usepackage[capitalise]{cleveref}
\usepackage{svg}
\usepackage{makecell}
\usepackage{subcaption}
\usepackage{graphicx}
\usepackage{xcolor}
\usepackage{tablefootnote}

\newcommand{\gcotwoe}{\emph{gCO\textsubscript{2}e} }
\makeatletter
\newcommand\footnoteref[1]{\protected@xdef\@thefnmark{\ref{#1}}\@footnotemark}
\makeatother

\AtBeginDocument{%
  \providecommand\BibTeX{{%
    \normalfont B\kern-0.5em{\scshape i\kern-0.25em b}\kern-0.8em\TeX}}}

\copyrightyear{2025}
\acmYear{2025}
\setcopyright{rightsretained}
\acmConference[]{}{}
\acmBooktitle{}\acmDOI{}
\acmISBN{}

\begin{document}

\title[Green Recommender Systems]{Green Recommender Systems: Understanding and Minimizing the Carbon Footprint of AI-Powered Personalization}

\author{Lukas Wegmeth}
\authornotemark[1]
\email{lukas.wegmeth@uni-siegen.de}
\orcid{0000-0001-8848-9434}
\affiliation{%
  \institution{Intelligent Systems Group, University of Siegen}
  \city{Siegen}
  \country{Germany}
}
\author{Tobias Vente}
\authornote{Both authors contributed equally to this research.}
\email{tobias.vente@uni-siegen.de}
\orcid{0009-0003-8881-2379}
\affiliation{%
  \institution{Intelligent Systems Group, University of Siegen}
  \city{Siegen}
  \country{Germany}
}
\author{Alan Said}
\email{alan@gu.se}
\orcid{0000-0002-2929-0529}
\affiliation{%
  \institution{University of Gothenburg}
  \city{Gothenburg}
  \country{Sweden}
}
\author{Joeran Beel}
\email{joeran.beel@uni-siegen.de}
\orcid{0000-0002-4537-5573}
\affiliation{%
  \institution{Intelligent Systems Group, University of Siegen}
  \city{Siegen}
  \country{Germany}
}

\renewcommand{\shortauthors}{Wegmeth and Vente, et al.}

\ccsdesc[500]{Information systems~Recommender systems}
\ccsdesc[500]{Hardware~Impact on the environment}

\keywords{Recommender Systems, Reproducibility, Carbon Footprint, Deep Learning, Energy Consumption, Green Computing, Green Recommender Systems}

\input{content/abstract}

\maketitle

\input{content/introduction}
\input{content/related_work}
\input{content/paper_study}
\input{content/method}
\input{content/results}
\input{content/discussion}
\input{content/guidelines}

\input{content/conclusions}

\bibliographystyle{ACM-Reference-Format}
\bibliography{sample-base}

\input{content/appendix}

\end{document}

%% file: content/abstract.tex
\begin{abstract}
As global warming soars, the need to assess and reduce the environmental impact of recommender systems is becoming increasingly urgent. 
Despite this, the recommender systems community hardly understands, addresses, and evaluates the environmental impact of their work. 
In this study, we examine the environmental impact of recommender systems research by reproducing typical experimental pipelines. 
Based on our results, we provide guidelines for researchers and practitioners on how to minimize the environmental footprint of their work and implement green recommender systems --- recommender systems designed to minimize their energy consumption and carbon footprint.
Our analysis covers 79 papers from the 2013 and 2023 ACM RecSys conferences, comparing traditional ``good old-fashioned AI'' models with modern deep learning models. 
We designed and reproduced representative experimental pipelines for both years, measuring energy consumption using a hardware energy meter and converting it into CO\textsubscript{2} equivalents. 
Our results show that papers utilizing deep learning models emit approximately 42 times more CO\textsubscript{2} equivalents than papers using traditional models. 
On average, a single deep learning-based paper generates 2,909 kilograms of CO\textsubscript{2} equivalents --- more than the carbon emissions of a person flying from New York City to Melbourne or the amount of CO\textsubscript{2} sequestered by one tree over 260 years.
This work underscores the urgent need for the recommender systems and wider machine learning communities to adopt green AI principles, balancing algorithmic advancements and environmental responsibility to build a sustainable future with AI-powered personalization.
\end{abstract}

%% file: content/introduction.tex
\section{Introduction}\label{introduction}
Carbon emissions play a pivotal role in global warming by driving the greenhouse effect that leads to rising temperatures and extreme weather events \cite{nukusheva2021global, kweku2018greenhouse, mitchell1989greenhouse, hansen2013assessing, sinha2019review}. 
With the ambitious goal of the \emph{United Nations Framework Convention on Climate Change} to cap the global warming temperature increase at 1.5 degrees Celsius\footnote{\url{https://unfccc.int/documents/184656}}, reducing carbon emissions becomes crucial \cite{hansen2013assessing}.
Therefore, significant reductions in carbon emissions are urgently needed to meet the temperature target and mitigate impact on the climate \cite{hansen2013assessing, sinha2019review, yoro2020co2}.

Concurrently, computationally heavy algorithms have become the norm for modern recommender systems, increasing the energy consumption of recommender systems experiments \cite{Chen_Zhang_Zhang_Dai_Yi_Zhang_Zhang_2020}. 
This trend is driven by a shift from traditional models like \emph{ItemKNN} (or so-called \emph{good old-fashioned AI}) to more sophisticated deep learning techniques \cite{10.1145/3285029, 8529185}.
The increased energy consumption of training deep learning models results in higher carbon emissions \cite{pmlr-v119-li20m}, further exacerbating environmental challenges.
With this backdrop, a relevant question is \emph{what is the environmental impact of recommender systems experiments?}

Despite its importance, only a few recent studies have addressed the energy consumption or the environmental impact of recommender systems \cite{10.1145/3604915.3608840,10.1007/978-3-031-87654-7_7,10.1007/978-3-031-87654-7_11,10.1007/978-3-031-87654-7_12,10.1007/978-3-031-87654-7_3,10.1007/978-3-031-87654-7_8,10.1007/978-3-031-87654-7_10,10.1007/978-3-031-87654-7_9}, with one formally introducing the concept of ``green recommender systems'' \cite{10.1145/3640457.3688160}. 
Given the lack of a clear definition, we define green recommender systems as systems designed to minimize their environmental impact across their life cycle, from research and development to deployment and operation.
These systems aim to match the performance of traditional models while prioritizing sustainability, even at the cost of trade-offs in metrics like accuracy. 
Achieving this goal requires reducing energy consumption and carbon emissions at every stage of the recommender systems project \cite{beel_said_vente_wegmeth_2024}.

The recommender systems community, comprising researchers and practitioners working on recommender systems, has largely overlooked one of the foundational aspects of green recommender systems: understanding the energy consumption and carbon emissions associated with their experiments \cite{10.1145/3604915.3608840}. 
This neglect is particularly troubling given the growing urgency to address carbon emissions globally \cite{1130574323982454028}.

\phantomsection
\label{para:researchQuestions} 
Therefore, our goal is to answer the question: What are the ecological costs of recommender systems research, past and present?
In this paper, we answer the following research questions:

\begin{enumerate}
    \item[\textbf{RQ1:}] How large is the energy consumption of a modern recommender systems research paper?
    \item[\textbf{RQ2:}] How substantial is the energy consumption and performance trade-off between traditional and deep learning models in recommender system experiments?
    \item[\textbf{RQ3:}] How has the carbon footprint of recommender systems experiments changed with the transition from traditional to deep learning models?
\end{enumerate}

In this work, we reproduce \emph{representative} recommender systems experimental pipelines and show the extent of carbon emissions attributable to recommender systems experiments, providing a comparative analysis of recommender systems experiments a decade apart.
Our analysis is based on 79 full papers from the 2013 and 2023 ACM Conferences on Recommender Systems\footnote{\url{http://recsys.acm.org}} (ACM RecSys). 
Reproducing a representative experiment setup for each year lets us measure the energy consumption of training a model and predicting item ranking lists. 
This approach also helps us evaluate their carbon footprint across different hardware configurations, including laptops, workstations, and desktop PCs.

Our contribution is a comprehensive analysis of the carbon emissions associated with recommender system experiments and the reproduction of a representative recommender systems pipeline comparing 13 datasets and 23 models from 2013 and 2023 on four hardware configurations. 
Our results indicate that a recommender systems research paper using deep learning models produces, on average, 3,297 kilograms of carbon dioxide (CO\textsubscript{2}) equivalents.
A research paper from 2023 emits approximately 42 times more CO\textsubscript{2} equivalents than a paper from 2013.
Furthermore, we demonstrate that the geographical location, which is correlated with the methods of energy production, such as fossil fuels or renewable sources, can result in up to a 12-fold variation in the carbon footprint of recommender systems research experiments.
Additionally, we found vast differences in the energy efficiency of different hardware configurations when performing recommender systems experiments, changing the carbon footprint of the same experiments by a factor of up to 10.
Based on our results, we provide practical guidelines to the recommender systems community with actionable strategies to minimize the carbon footprint of recommender systems and develop green recommender systems.
This manuscript significantly extends our previous work that was published at ACM RecSys 2024 \cite{10.1145/3640457.3688074}. 

%% file: content/related_work.tex
\section{Related Work} 
Greenhouse gases, such as CO\textsubscript{2}, trap heat in the atmosphere and drive the greenhouse effect \cite{IPCC2021Glossary}. 
Their environmental impacts have been extensively studied, revealing far-reaching consequences including rising global temperatures \cite{Kweku2018GreenhouseEG}, accelerated polar ice cap melting \cite{Cao2017EnhancedWG}, rising sea levels \cite{Barth1988GreenhouseEA}, and an increase in extreme weather events \cite{Anglil2014AttributionOE,Paik2020DeterminingTA,Touma2021HumandrivenGG}.
These effects threaten ecosystems and biodiversity and pose significant challenges to human societies, including food security \cite{Schmidhuber2007GlobalFS}, health risks \cite{Hayes2018ClimateCA,Tong2019PreventingAM}, and economic instability \cite{Tol2009TheEE}.
The scientific consensus underscores the anthropogenic nature of climate change, linking it directly to human activities such as deforestation, industrialization, and energy production \cite{Lynas2021, Cook2016}.
The urgency of addressing greenhouse gas emissions is underscored by the goal of limiting global warming to 1.5°C above pre-industrial levels, as outlined in the Paris Agreement, which requires rapid and substantial reductions in emissions across all sectors, including energy-intensive industries like computing \cite{ipcc_sr15_2018}.

Among the greenhouse gases, CO\textsubscript{2} \cite{Cain2019} is the most prevalent \cite{climatewatch_ghg_2022}, emitted primarily from fossil fuel combustion for energy generation and agricultural and industrial processes \cite{iea_co2_2023}.
Efforts are being made to document the development of greenhouse gas emissions globally \cite{crippa_ghg_2024}.
Furthermore, energy carries a direct monetary value and a cost associated with the damage caused by greenhouse gas emissions from its production, commonly referred to as \emph{social cost} \cite{Ricke2018}. 
With technological advancements driving increasing energy demand, computing has emerged as a significant contributor to global emissions \cite{9407142}, necessitating a closer examination of its environmental footprint. 
While the specific share of recommender systems within the broader computing footprint remains unclear, their growing adoption suggests that their energy demands merit closer scrutiny in pursuing more sustainable practices.

The growing demand for computational power has led to significant concerns about the environmental impact of computing, e.g., in cryptocurrency mining \cite{Khler2019LifeCA,Chao2020EnergyCO,Wendl2022TheEI} and modern machine learning applications \cite{Kaack2022AligningAI,Wu2021SustainableAE}.
Growing awareness of the carbon footprint of machine learning has driven the creation of standards for measuring, e.g., via direct hardware measurement \cite{GARCIAMARTIN201975,10603302}, and minimizing energy consumption \cite{Yigitcanlar2021GreenAI,Lacoste2019QuantifyingTC,Luccioni2023CountingCA} during model training and deployment.
Several studies have highlighted the need for improved energy efficiency, with particular emphasis on large-scale neural networks, e.g., in natural language processing \cite{strubell2019energy}, in computer vision \cite{https://doi.org/10.48550/arxiv.2311.00447}, and in large language models \cite{Luccioni2022EstimatingTC,Faiz2023LLMCarbonMT,wu2023survey}, which are notorious for their high computational costs. 
As a result, there has been an increasing focus on developing energy-efficient models, optimizing hardware, and exploring greener alternatives to conventional training processes \cite{Salehi2024DataCentricGA,Verdecchia2023ASR,10214579,budennyy2022eco2ai,mehta2023review,van2021sustainable,Lannelongue2023,betello2025searchfitsallparetooptimal}. 
For instance, research has proposed more efficient techniques for hyperparameter optimization \cite{pmlr-v80-falkner18a}, model pruning \cite{tmamna24}, and transfer learning \cite{DBLP:journals/corr/abs-2104-10350} to reduce the carbon footprint of machine learning models.
Libraries and frameworks have emerged to help researchers measure and track the energy consumption of their experiments, offering tools to quantify the carbon footprint of machine learning workflows \cite{Anthony2020CarbontrackerTA,JMLR:v21:20-312,powermeter}. 
Despite these, the intersection of recommender systems and green computing remains underexplored.
While it is well-established that energy consumption is a key concern in natural language processing and computer vision, recommender systems have not yet been subject to the same level of scrutiny concerning their environmental impact.

The sustainability of recommender systems is only recently gaining attention in the recommender systems research community, with all relevant works published within the last two years. 
Spillo et al. \cite{10.1145/3604915.3608840} are the first to examine the trade-off between recommendation accuracy and carbon emissions, advocating for more energy-efficient models.
Their subsequent work \cite{10.1145/3640457.3688160} further explores how dataset reduction techniques can balance performance and environmental impact.
The growing interest in this field led to the ``First International Workshop on Recommender Systems for Sustainability and Social Good''\footnote{\url{https://recsogood.github.io/recsogood24/}} (RecSoGood) at ACM RecSys 2024, which specifically addresses sustainable recommender systems.
Furthermore, the workshop, along with similar initiatives in related fields\footnote{\url{https://www.ecir2024.org/ir4good/}, \url{https://www2025.thewebconf.org/web4good}} have spurred an interest in making recommender systems follow a path for ``recommender systems for good'' \cite{jannach_recommender_2024,said_recommender_2024}.

Several key studies emerged from the RecSoGood workshop: Arabzadeh et al. \cite{10.1007/978-3-031-87654-7_7} propose methods to optimize dataset size while maintaining predictive performance.
Purificato et al. investigate \cite{10.1007/978-3-031-87654-7_11} the environmental impact of graph neural network-based recommenders by comparing model efficiency.
Similarly, Plaza et al. \cite{10.1007/978-3-031-87654-7_12} analyze the carbon footprint of session-based recommenders through empirical energy measurements.
Baumgart et al. \cite{10.1007/978-3-031-87654-7_9} introduce `e'-fold cross-validation, an energy-efficient alternative to `k'-fold validation, significantly reducing the energy consumption of recommender systems cross-validation.
Banerjee et al. \cite{10.1007/978-3-031-87654-7_3} develop a sustainability-aware metric for tourism recommendations.
While these works address specific aspects of recommender system sustainability, none provide a comprehensive analysis of the carbon footprint across the entire research life cycle. 
This work fills that gap by offering a holistic assessment of the environmental impact of recommender systems research.

Most existing works that directly measure the energy consumption of recommender systems rely on CodeCarbon \cite{benoit_courty_2024_11171501}, which measures energy consumption through hardware sensors of the computing resource or approximates energy consumption based on historic data.
To address this limitation, Wegmeth et al. \cite{10.1007/978-3-031-87654-7_8} release EMERS, a software tool for recommender systems that enables accurate, hardware-based energy measurements of entire computing resources through a smart plug.
Alternatively, Spillo et al. \cite{10.1007/978-3-031-87654-7_10} present a regression model that estimates the energy consumption of recommender systems models.
Recommender systems are also explored for their potential to enhance sustainability and energy efficiency in other domains \cite{HIMEUR20211,af69af4ea12246b6b0c0d73908ed1fc5,zhou2024advancingsustainabilityrecommendersystems}.
Furthermore, research on automated recommender systems considers computing power requirements \cite{Wegmeth2022-bx,vente2023introducing}, significantly impacting energy consumption.
However, accurately estimating the global carbon footprint of recommender systems remains impossible without further research.
This gap in research highlights the importance of further exploration into the environmental footprint of recommender systems, which, despite their pervasive use, have yet to receive the same level of scrutiny as other machine-learning domains.

%% file: content/paper_study.tex
\section{Comparative Study: Recommender Systems Research in 2023 versus 2013}\label{paper_review}
We summarize the historical development of recommender systems experiments by analyzing research papers from 2023 and 2013, reflected through peer-reviewed papers at ACM RecSys.
This approach allows us to reproduce measurements and estimations to answer our research questions.
To this end, we present our analysis of all \emph{full papers} accepted in the main track at ACM RecSys in 2013 (32 papers~\cite{10.1145/2507157}) and 2023 (47 papers~\cite{10.1145/3604915}).
All papers considered in this analysis are published in the ACM Digital Library.

In the following paragraphs, we examine these recommender systems papers for \textbf{hardware} specifications, software \textbf{libraries}, design decisions in \textbf{experimental pipelines}, the availability of \textbf{open-source code} for reproducibility, and used \textbf{datasets}. 
\cref{study_table} provides an overview by summarizing the comparisons.

\begin{table}[h!]
\centering
\caption{Comparison of recommender systems research papers from ACM RecSys 2013 and 2023.}
\label{tab:comparison}
\begin{tabular}{l|l|l}
\toprule
\textbf{Statistic}                    & \textbf{ACM RecSys 2013}                   & \textbf{ACM RecSys 2023}                     \\ \midrule
Number of accepted papers                      & 32                              & 47                                \\
Papers detailing hardware             & 6 papers (19\%)      & 15 papers (32\%)          \\
Most common hardware & Intel Xeon CPU & NVIDIA V100 GPU \\
Most common libraries                 & MyMediaLite, Mahout, InferNet   & PyTorch, TensorFlow, RecBole      \\
Data splitting technique              & Holdout (20 papers, 63\%)       & Holdout (20 papers, 42\%)         \\
Hyperparameter optimization           & 12 papers (38\%)                & 34 papers (72\%)                  \\
Most common optimization technique    & Grid search (12 papers, 38\%)   & Grid search (31 papers, 66\%)     \\
Most common evaluation metric                    & Precision (13 papers, 40\%)     & nDCG (32 papers, 68\%)            \\
Most common prediction task                      & Rating prediction (56\%)        & Top-n ranking prediction (94\%)   \\
Papers sharing code                   & 1 paper (3\%)                   & 29 papers (62\%)                  \\
Mean datasets used in each paper               & 2.19                            & 2.85                              \\
Most popular datasets                 & MovieLens, LastFM, Netflix      & Amazon, MovieLens, LastFM  \\
\bottomrule
\end{tabular}
\label{study_table}
\end{table}

\paragraph{Hardware:} 
Only 15 (32\%) full papers from ACM RecSys 2023 detail the hardware used for experiments, and consequently, 32 (68\%) of the papers do not report this information.
All these 15 papers explicitly report the usage of NVIDIA GPUs, with seven specifically mentioning the NVIDIA V100 GPU (released in 2017). 

Looking back at papers published at ACM RecSys 2013, only 6 out of 32 (19\%) contain information about the hardware used, meaning 26 (81\%) do not contain hardware information.
Contrary to 2023, none of the papers from 2013 mention using a GPU. 
However, all 6 papers that disclose their hardware utilize an Intel Xeon CPU.

\paragraph{Libraries:} 
At ACM RecSys 2023, 19 (40\%) papers use PyTorch, and 6 (13\%) papers use TensorFlow, implementing deep learning recommender systems.
RecBole \cite{recbole[1.2.0]} is used by 5 (11\%) papers, making it the most popular library designed explicitly for recommender systems. 
While 12 (26\%) papers from 2023 neither specify the library used nor provide accessible code\footnote{The code was either not made publicly available (10 papers) or became inaccessible due to broken links (2 papers).}, they all implement deep learning models.
The remaining 5 (10\%) papers report using one of the following libraries: Elliot \cite{DBLP:conf/sigir/AnelliBFMMPDN21}, ReChorus \cite{wang2020make}, Bambi \cite{Capretto2022}, FuxiCTR \cite{DBLP:conf/sigir/ZhuDSMLCXZ22}, or CSRLab \cite{crslab}.

This contrasts the patterns observed in ACM RecSys 2013 papers. 
The majority of papers, 27 (84\%), do not report using any open libraries. 
Instead, they rely on custom model implementations.
Only 5 (16\%) papers report using libraries, which include MyMediaLite \cite{Gantner2011MyMediaLite}, Apache Mahout \cite{10.5555/3455716.3455843}, InferNet \cite{InferNET18}, and libpMF \cite{hfy12a}. 
This shift highlights a considerable evolution in adopting standardized libraries within recommender systems over the decade between 2013 and 2023. 
Arguably, the feasibility of reproducing a recommender systems paper has increased over time, at least in terms of software.

\paragraph{Experimental Pipeline:} 
The experimental pipeline remains consistent between papers from 2023 and 2013, but there are significant differences in the models and datasets.
For instance, the holdout split is the most popular data splitting technique, used in 20 (42\%) of the 2023 papers and 20 (63\%) of the 2013 papers.
Additionally, 22 (47\%) papers in 2023 and 20 (63\%) in 2013 do not employ dataset pruning, although n-core pruning appears to have gained notable popularity, appearing in 19 (40\%) papers in 2023.

One of the most significant differences lies in the evaluation metrics used: nDCG is the predominant metric in 2023, used in 32 (68\%) papers, whereas, in 2013, Precision is the most popular one, appearing in 13 (40\%) papers, followed by RMSE and nDCG, each used in 5 (16\%) papers.
This is partly because 44 (94\%) of papers in 2023 focus on top-n ranking prediction tasks, while 18 (56\%) of papers in 2013 focus on rating prediction tasks.

Furthermore, grid search is the favored optimization technique in 2023, used in 31 (66\%) papers, while 13 (28\%) papers do not report optimizing hyperparameters.
The remaining 3 (6\%) papers apply other techniques, e.g., Bayesian optimization or random search.
In contrast, only 1 (3\%) paper in 2013 explicitly applied grid search, while 20 (63\%) reported no hyperparameter optimization, e.g., they used hyperparameters from other works or set hyperparameters without disclosing their search procedure.
However, while the remaining 11 (34\%) papers do not specify a hyperparameter optimization technique, they describe an optimization process akin to grid search.

\paragraph{Open-Source Code:}
Only 18 (38\%) of ACM RecSys 2023 papers do not contain links to their source code. 
On the other hand, 29 (62\%) make their code available, with all but 2 hosting it on \emph{GitHub} and the remaining hosting it on their organization's website.
However, 3 (6\%) repositories linked in these papers are empty or unreachable.
Only 1 (3\%) paper from ACM RecSys 2013 shares code. 

\paragraph{Datasets:} 
On average, papers from ACM RecSys 2023 include three datasets (mean of 2.85). 
The most frequent datasets are from the Amazon2018 series, appearing in 15 (32\%) papers, followed by the MovieLens datasets in 13 (28\%), of which MovieLens-1M is used in 11 (23\%) papers, while MovieLens-100K, MovieLens-10M, and MovieLens-20M are used in 2 (4\%) papers. 
Other commonly used datasets are LastFM in 6 (13\%) papers, Yelp-2018 in 5 (11\%) papers, and Gowalla in 5 (11\%) papers.

In contrast to 2023, papers at ACM RecSys 2013 include, on average, two distinct datasets in their experimental pipeline (mean of 2.19). 
The most frequently used datasets are the MovieLens datasets, used in 10 (31\%) papers. 
Neither of the popular Amazon datasets were available in 2013. 
Other commonly used datasets include the LastFM dataset, used in 4 (13\%) papers, and the Netflix Prize dataset, used in 3 (9\%) papers. 

%% file: content/method.tex
\section{Method}
We measure the energy consumption of 23 recommender system models applied to 13 datasets using a smart power plug.
To assess the impact of hardware on energy efficiency, we run the experiments on five distinct computers.
We estimate the carbon footprint by assessing the carbon emissions related to energy generation in five locations. 
Our experimental pipeline is based on the study in \cref{paper_review}, ensuring representative data collection. 
All design decisions are derived from this study unless otherwise specified.
The code used to execute and measure the experiments is publicly available in our GitHub\footnote{\url{https://github.com/ISG-Siegen/recsys-carbon-footprint}} repository and further contains documentation to ensure the reproducibility of our experiments.

\subsection{Experimental Pipeline}
We randomly divide each of the 13 datasets into three splits, where 60\% of the data is for training, 20\% for validation, and 20\% for testing. 
Since we focus on energy consumption rather than maximizing performance or generalizability, we do not employ cross-validation or repeat experiments.
Although this decision comes at the cost of reliability and performance in terms of accuracy, our goal is not to optimize the recommender models to beat a baseline but to measure the energy consumption of a representative recommender system experiment. 
We measure performance using nDCG@10 for top-n ranking predictions and RMSE for rating predictions.
To test the impact of hyperparameter configurations on energy consumption, we evaluated six different settings for three models from representative categories: one nearest neighbors approach, one matrix factorization method, and one deep learning model, across three datasets. 
However, for our final evaluation, we used the default hyperparameter settings provided by the libraries rather than optimized configurations. 
We made this decision to minimize unnecessary energy consumption.
All deep learning models are trained for 200 epochs, with model validation after every fifth epoch to facilitate early stopping.

\subsubsection{Datasets:}
We run our experiments on the 13 most representative datasets from the years 2013 and 2023, according to our paper study (\cref{paper_review}).
For top-n prediction tasks, we convert rating prediction datasets according to the practice that is most commonly found in our paper study \cite{10.1145/3604915.3608802,10.1145/3604915.3608791,10.1145/3604915.3608771,10.1145/3604915.3608785}.
Furthermore, we prune all datasets such that all included users and items have at least five interactions, commonly known as five-core pruning \cite{10.1145/3357384.3357895,10.1145/3460231.3474275,10.1145/3523227.3546770}.
\cref{tab:dataset_statistics} shows the dataset statistics of all included datasets for the preprocessed datasets.

\renewcommand{\arraystretch}{1.2}
\begin{table}
\caption{Basic information of the data sets used in our experiments after preprocessing.}
\begin{tabular}{l|r|r|r|r}
\toprule
\textbf{Dataset Name} & \textbf{Users} & \textbf{Items} & \textbf{Interactions} & \textbf{Density} \\
\midrule
\makecell[tl]{Amazon2018-Books \cite{ni-etal-2019-justifying}} & 105,436 & 151,802 & 1,724,703 & 0.0108 \\
\makecell[tl]{Amazon2018-CDs-And-Vinyl \cite{ni-etal-2019-justifying}} & 71,943 & 107,546 & 1,377,008 & 0.0178 \\
\makecell[tl]{Amazon2018-Electronics \cite{ni-etal-2019-justifying}} & 62,617 & 187,288 & 1,476,535 & 0.0126 \\
\makecell[tl]{Amazon2018-Sports-And-Outdoors \cite{ni-etal-2019-justifying}} & 69,781 & 185,024 & 1,498,609 & 0.0116 \\
\makecell[tl]{Amazon2018-Toys-And-Games \cite{ni-etal-2019-justifying}} & 75,856 & 192,326 & 1,686,250 & 0.0116 \\
Gowalla \cite{10.1145/2020408.2020579} & 64,115 & 164,532 & 2,018,421 & 0.0191 \\
Hetrec-LastFM \cite{Cantador:RecSys2011} & 1,090 & 3,646 & 52,551 & 1.3223 \\
\makecell[tl]{MovieLens-100K \cite{10.1145/2827872}} & 943 & 1,349 & 99,287 & 7.8049 \\
\makecell[tl]{MovieLens-1M \cite{10.1145/2827872}} & 6,040 & 3,416 & 999,611 & 4.8448 \\
\makecell[tl]{MovieLens-Latest-Small \cite{10.1145/2827872}} & 610 & 3,650 & 90,274 & 4.0545 \\
Netflix\tablefootnote{\url{https://www.kaggle.com/datasets/netflix-inc/netflix-prize-data}} & 11,927 & 11,934 & 5,850,559 & 4.1103 \\
Retailrocket\tablefootnote{\url{https://www.kaggle.com/datasets/retailrocket/ecommerce-dataset}} & 22,178 & 17,803 & 240,938 & 0.0610 \\
Yelp-2018\tablefootnote{\url{https://www.yelp.com/dataset}} & 213,170 & 94,304 & 3,277,931 & 0.0163 \\
\bottomrule
\end{tabular}
\label{tab:dataset_statistics}
\end{table}
\renewcommand{\arraystretch}{1.0}

\subsubsection{Models:}
We include the 23 most frequently used models from the years 2013 and 2023 in our experiments, according to our paper study (\cref{paper_review}).
The models are from RecBole \cite{recbole[1.2.0]} (indicated by \emph{RB}), RecPack \cite{10.1145/3523227.3551472}  (indicated by \emph{RP}) and LensKit \cite{ekstrand2020lenskit} (indicated by \emph{LK}).
We run models from RecBole on a GPU while we run models from LensKit and RecPack on a CPU. 
\cref{tab:algorithms} shows all models used in our experiments.

\begin{table}
\caption{Models used in our experiments.}
\begin{tabular}{l|l|l}
\toprule
\textbf{Library} & \textbf{Executed on} & \textbf{Model} \\
\midrule
RecBole & GPU & \makecell[tl]{BPR, DGCF, DMF, ItemKNN, LightGCN, MacridVAE, MultiVAE, \\NAIS, NCL, NeuMF, NGCF, Popularity, RecVAE, SGL}  \\
\midrule
RecPack & CPU & \makecell[tl]{ItemKNN, NMF, SVD} \\
\midrule
LensKit & CPU & \makecell[tl]{ImplicitMF, ItemKNN, UserKNN, Popularity, BiasedMF, FunkSVD} \\
\midrule
Elliot & CPU & \makecell[tl]{AMF, BPR, Popularity} \\
\midrule
Elliot & GPU & \makecell[tl]{MultiVAE, MultiDAE} \\
\bottomrule
\end{tabular}
\label{tab:algorithms}
\end{table}

\subsection{Representative Pipelines}
\label{rep_pipeline}
Our research paper study indicates that experimental pipelines from 2013 and 2023 exhibit notable differences (\cref{paper_review}). 
For instance, in 2023, all but three papers focused on top-n ranking prediction tasks, whereas in 2013, around half focused on rating prediction tasks. 
Additionally, in 2023, experiments often utilize datasets from the \emph{Amazon-2018} series, which did not exist in 2013. 
Consequently, we introduce three distinct representative pipelines to account for these differences.
\cref{tab:specific_pipelines} provides an overview of the models and datasets used for specific pipelines. 

\begin{table}
\caption{Representative pipelines used to run experiments.}  
\begin{tabular}{l|l|l}
\toprule
\makecell[tl]{\textbf{Year and}\\ \textbf{Prediction Type}} & \textbf{Models} & \textbf{Datasets} \\
\midrule
\makecell[tl]{2013\\Rating\\Prediction} & \makecell[tl]{ItemKNN$^{LK}$, UserKNN$^{LK}$, BiasedMF$^{LK}$,\\ FunkSVD$^{LK}$} & \makecell[tl]{Movielens-100K, Movielens-1M,\\ Netflix} \\
\midrule
\makecell[tl]{2013\\Top-N\\Ranking\\Prediction} & \makecell[tl]{ImplicitMF$^{LK}$, ItemKNN$^{LK}$, UserKNN$^{LK}$,\\ Popularity$^{LK}$, ItemKNN$^{RP}$, NMF$^{RP}$, SVD$^{RP}$,\\ BPR$^{RB}$, ItemKNN$^{RB}$, Popularity$^{RB}$} & \makecell[tl]{Hetrec-LastFM, Movielens-100K,\\ Movielens-1M, Gowalla} \\  
\midrule
\makecell[tl]{2023\\Top-N\\Ranking\\Prediction} & 
\makecell[tl]{ImplicitMF$^{LK}$, ItemKNN$^{LK}$, UserKNN$^{LK}$,\\Popularity$^{LK}$, ItemKNN$^{RP}$, NMF$^{RP}$, SVD$^{RP}$,\\ BPR$^{RB}$, DGCF$^{RB}$, DMF$^{RB}$, ItemKNN$^{RB}$,\\ LightGCN$^{RB}$, MacridVAE$^{RB}$, MultiVAE$^{RB}$,\\ NAIS$^{RB}$, NCL$^{RB}$, NeuMF$^{RB}$, NGCF$^{RB}$,\\ Popularity$^{RB}$, RecVAE$^{RB}$, SGL$^{RB}$, AMF$^{EL}$,\\ BPR$^{EL}$, MultiDAE$^{EL}$, MultiVAE$^{EL}$,\\ Popularity$^{EL}$} & \makecell[tl]{Gowalla, Hetrec-LastFM,\\ MovieLens-100K, MovieLens-1M,\\ MovieLens-Latest-Small,\\ Amazon2018-Electronics,\\ Amazon2018-Toys-And-Games,\\ Amazon2018-CDs-And-Vinyl,\\ Amazon2018-Sports-And-Outdoors,\\ Amazon2018-Books,\\ Yelp-2018, Retailrocket}\\
\bottomrule
\end{tabular}
\label{tab:specific_pipelines}
\end{table}

\subsection{Calculating Greenhouse Gas Emission}
To calculate the greenhouse gas emissions from recommender systems experiments, we first record the energy consumption and calculate the equivalent emissions based on the consumption. 

\subsubsection{Measuring Electrical Energy Consumption}
To accurately measure the energy consumption of recommender system experiments, we equip each computer in our setup with a commercially available smart power plug\footnote{\url{https://www.shelly.com/en-se/products/product-overview/shelly-plus-plug-s}}.
The plug provides status, diagnostics, and energy measurements through multiple interfaces, including a smart home application, a web interface, a cloud API, and an HTTP API.
We use the HTTP API for data collection, enabling fine-grained measurements of both instantaneous power draw in Watts and cumulative energy consumption in kilowatt-hours (kWh) at 500-millisecond intervals, which is the fastest measurement interval of the plug. 
We consider this sampling rate sufficiently precise to capture the start and end of training, prediction, and evaluation phases, which typically span seconds to hours, rendering potential half-second deviations negligible.
For additional accuracy, we interpolate energy consumption between the nearest two measurements around the beginning and end of each experimental phase.

We synchronize the clocks of the power plug and the computing resource, then measure the baseline idle power draw with no experiments or background applications running.
After finishing the experiments, we align the power plug's measurements, labeled with Unix timestamps, with the experiment logs using their corresponding Unix timestamps.
Our measurements account for the full hardware energy usage, including all components such as cooling, power supply, CPU, GPU, and memory, as well as the overhead from the operating system.

To account for potential variability from external factors, e.g., thermal throttling, power delivery fluctuations, we conducted three repetitions of all experiments that run across all included hardware configurations.

\subsubsection{Calculating Greenhouse Gas Emissions Based on Energy Consumption}\label{energy_to_carbon}
We convert the measured energy consumption in kWh into carbon dioxide equivalents (CO\textsubscript{2}e) utilizing the comprehensive dataset provided by \emph{Ember}\footnote{\url{https://ember-climate.org/}}. 
The Ember dataset aggregates data from the European Electricity Review and information from various original data providers into a singular dataset\footnote{\url{https://docs.owid.io/projects/etl/}}. 

The Ember dataset features a conversion rate from kWh to gCO\textsubscript{2}e. 
CO\textsubscript{2}e represents the greenhouse gas emissions released and the overall environmental impact of energy generation. 
Since the hardware does not directly impact the environment by, e.g., emitting greenhouse gases, we utilize the carbon dioxide equivalents associated with the energy generation process of the consumed energy.

The gCO\textsubscript{2}e conversion rate linked to energy generation varies notably based on the method of production \cite{kerem2022investigation}.
Energy from renewable sources, such as hydropower, typically has a lower carbon footprint than coal combustion (\cref{ghg}).
Different regions employ diverse energy generation methods, so we use the global average conversion rate and compare various geographical locations.
However, this is a macro view providing an average value, ignoring, e.g., the season of the year or weather changes. 

\subsection{Hardware}
We conduct all experiments across five computers with different hardware configurations from different years to assess the impact of hardware efficiency on energy consumption and, consequently, the carbon footprint.
Furthermore, we evaluate whether improvements in hardware efficiency can offset the energy demands of transitioning to deep learning. 
We present the hardware specifications used in our experiments in \cref{tab:hardware}.

\stepcounter{footnote}
\begin{table}
\caption{Hardware used to compare hardware efficiency.}
\begin{tabular}{l|l|l|l|l|l}
\toprule
\makecell[tl]{\textbf{Computer and}\\\textbf{Year}} & \textbf{CPU} & \textbf{GPU} & \textbf{RAM} & \textbf{Storage} & \textbf{Cooling}\\
\midrule
\makecell[tl]{Modern Workstation I\\2023} &  \makecell[tl]{Intel\\ Xeon W-2255\\ 10-Core @ 3.70GHz\\ TDP: 165W} &  \makecell[tl]{NVIDIA\\ GeForce RTX 3090\\ TDP: 350W} & 256GB & 2TB & Air \\
\midrule
\makecell[tl]{Modern Workstation II\\2023} & \makecell[tl]{AMD \\Ryzen Threadripper Pro 3975WX \\32-Core @ 3.5GHz\\ TDP: 280W} &  \makecell[tl]{NVIDIA\\ GeForce RTX 3090\\ TDP: 350W} & 512GB & 5TB & Liquid \\
\midrule
\makecell[tl]{Mac Studio\\2022} & \makecell[tl]{M1 Ultra\\TDP: 70W\footnotemark[\value{footnote}]} & \makecell[tl]{M1 Ultra\\TDP: 70W\footnotemark[\value{footnote}]} & 64GB & 1TB & Air \\
\midrule
\makecell[tl]{MacBook Pro\\2020} & \makecell[tl]{M1\\TDP: 14W\footnotemark[\value{footnote}]} & \makecell[tl]{M1\\TDP: 14W\footnotemark[\value{footnote}]} & 16GB & 1TB & Air \\
\midrule
\makecell[tl]{Legacy Workstation\\2013} &  \makecell[tl]{Intel\\ Core i7-6700K\\ 4-Core @ 4.00GHz\\ TDP: 91W} &  \makecell[tl]{NVIDIA\\ GeForce GTX 980 Ti\\ TDP: 350W} & 128GB & 1TB & Air \\
\bottomrule
\end{tabular}
\label{tab:hardware}
\end{table}
\footnotetext{Estimated TDP as Apple has not released actual values: \url{https://nanoreview.net/en/cpu-compare/apple-m1-ultra-vs-apple-m1}}

%% file: content/results.tex
\section{Results}
Our results demonstrate the energy consumption, the trade-offs between energy and performance, and the carbon footprint associated with the 2013 and 2023 ACM RecSys full paper experiments.

\subsection{The Energy Consumption of a 2023 Recommender Systems Research Paper} \label{RQ1}
\begin{figure*}
    \centering
    \begin{subfigure}[b]{0.47\textwidth}
         \resizebox{1\linewidth}{!}{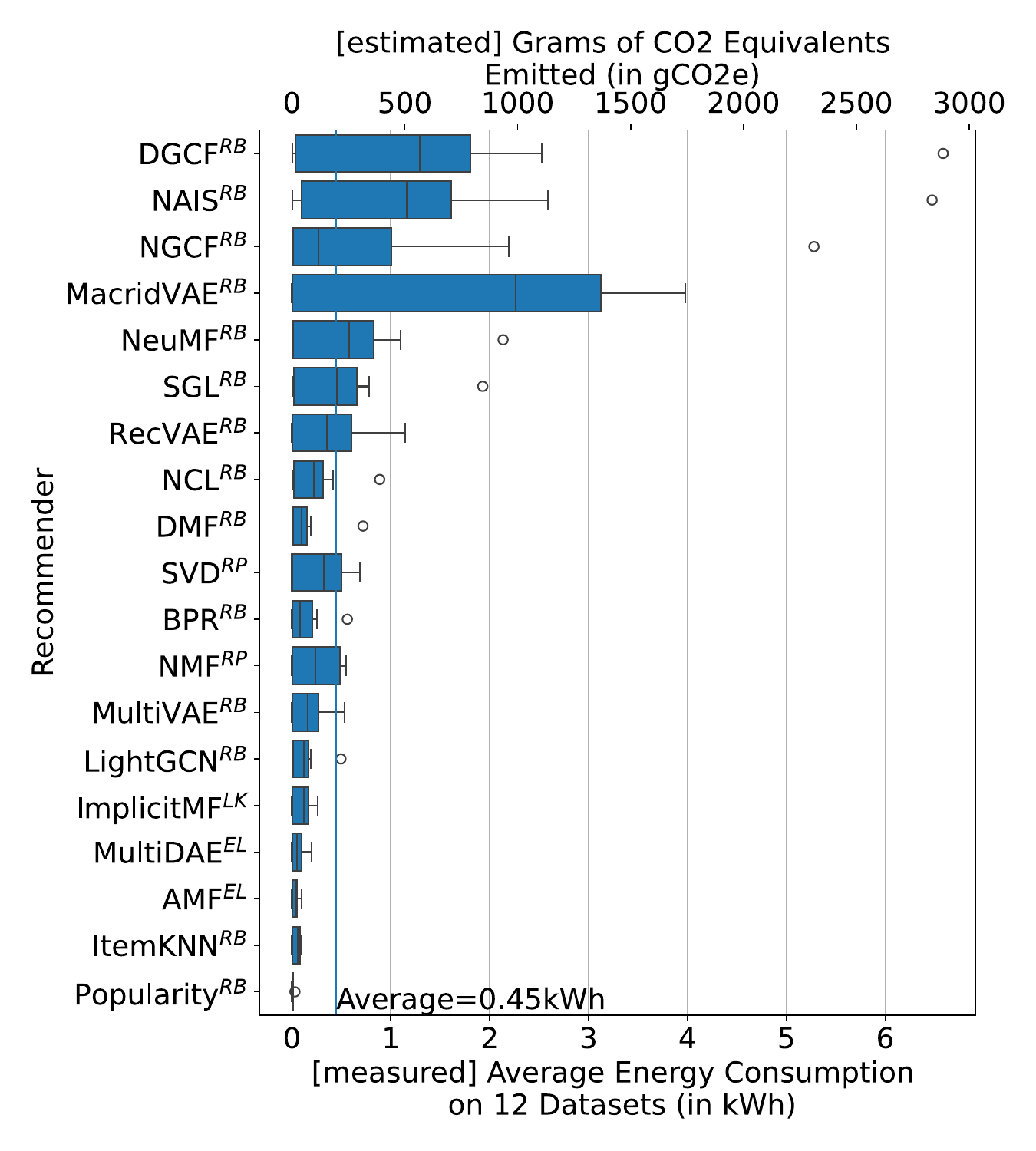}
         \caption{Energy Consumption for 19 Models}
         \label{fig:avgalgo}
    \end{subfigure}\hfill
    \begin{subfigure}[b]{0.53\textwidth}
        \resizebox{1\linewidth}{!}{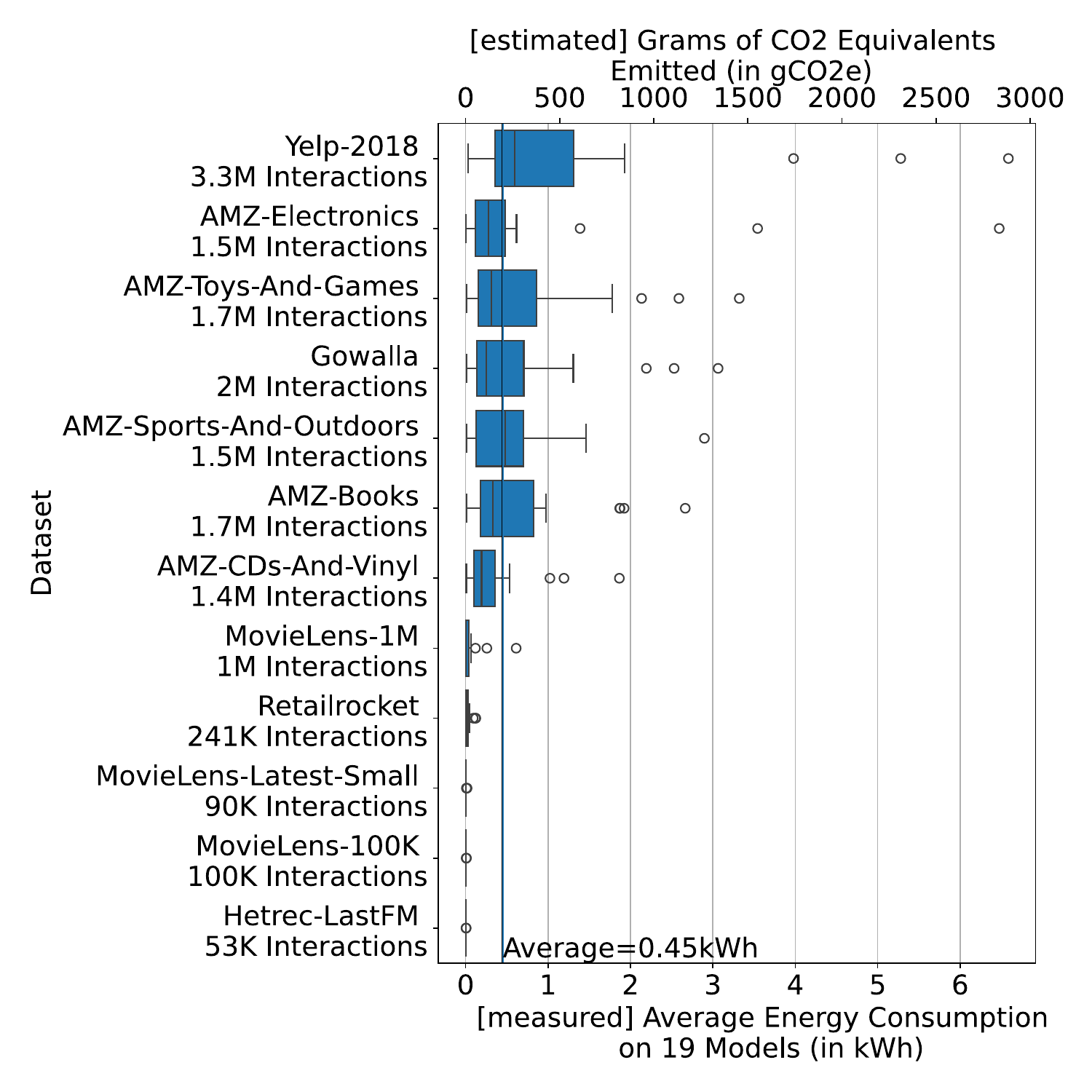}
        \caption{Energy Consumption for 12 Datasets}
        \label{fig:avgdata}
        
    \end{subfigure}\hfill
    \caption{Average energy consumption of recommender system models across twelve datasets, alongside the average consumption per dataset across nineteen models. \emph{Blue} vertical lines represent the average consumption in kWh. The upper x-axis displays the equivalent CO\textsubscript{2} emissions in grams (gCO\textsubscript{2}e), calculated based on the 2023 world average.}
    \Description[Test]{Test}
    \label{fig:avgpowcons}
\end{figure*}

\subsubsection{The Energy Consumption per Model Type and Dataset}\label{consumption_per_dataset}
Based on our experiments, we estimate the energy consumption of a single run of one recommender systems model on one dataset to be, on average\footnoteref{note1}, 0.45 kWh (\cref{fig:avgalgo}). 

Energy consumption varies among model types and datasets. 
Relatively simple models like \emph{Popularity} and \emph{ItemKNN} consume only 0.007 kWh and 0.04 kWh, respectively (average over twelve datasets used for experiments representing 2023\footnote{\label{note1}Throughout this work, reported values are based on the mean unless otherwise specified. However, boxplots depict the median, which may differ from the mean due to variations in data distribution.
Absolute numerical values and additional statistics for the data shown in \cref{fig:avgpowcons} are provided in the comprehensive tables in the \cref{ect}}).
Specific recent deep learning models consume a relatively low amount of energy, e.g., \emph{DMF} and \emph{LightCGN} consume 0.13 kWh and 0.12 kWh, respectively (average over twelve datasets used for 2023 experiments\footnoteref{note1}). 
Contrasting, the most ``expensive'' models -- \emph{MacridVAE} and \emph{DGCF} consume, on average\footnoteref{note1}, 1.79 kWh and 1.45 kWh, respectively. 
Consequently, the most expensive model (MacridVAE), regarding energy consumption, requires 257 times as much energy as the cheapest model (Popularity). 

The energy consumption of individual models on different datasets is even higher.
\emph{Popularity} consumes only 0.000036 kWh on \emph{Movielens-Latest-Small} but 0.03 kWh on the \emph{Yelp-2018} dataset (factor 800). 
The deep learning model \emph{DGCF} consumes 0.005 kWh on \emph{Hetrec-LastFM} but 6.6 kWh on \emph{Yelp-2018} (factor 1,444). 

When executing models over larger datasets, energy consumption increases compared to executing the same model on smaller data.
For example, the average recommender systems model consumes 0.001 kWh on \emph{Hetrec-LastFM} with 53 thousand interactions and 1.36 kWh, 1,360 times more energy, on \emph{Yelp-2018} with 3.3 million interactions (\cref{fig:avgdata}\footnoteref{note1}). 

However, the energy consumption per dataset is not solely dependent on the number of interactions. 
Although \emph{Amazon2018-Electronics} has \textasciitilde25\% fewer interactions than \emph{Gowalla} (1.5M vs. 2M; \cref{fig:avgdata}), models running on \emph{Amazon2018-Electronics} consume, on average\footnoteref{note1}, 19\% more energy (0.81 kWh/0.68 kWh). 
Similarly, even though \emph{Movielens-1M} includes around half the interactions of \emph{Gowalla} (1M vs. 2M), models on \emph{Movielens-1M} consume only 8\% of the energy \emph{Gowalla} needs (0.06 kWh vs. 0.68 kWh).

\subsubsection{Energy Impact of Training vs. Prediction}
Our experiments show that the energy consumption of recommender system models varies by dataset, model type, and operational stage, i.e., training vs. ranking prediction.

\begin{figure}
    \centering
    \resizebox{0.75\linewidth}{!}{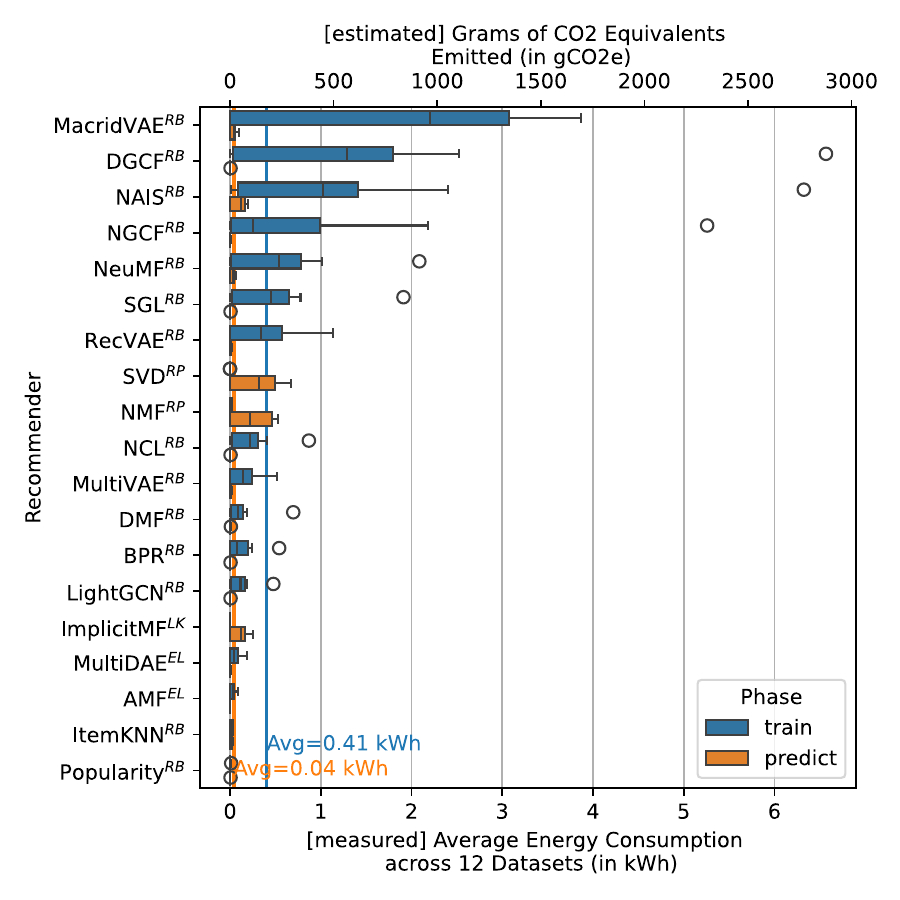}
    \caption{Average energy consumption of recommender system models for training and prediction across twelve datasets. \emph{Blue} vertical lines represent the average consumption of the training and the \emph{orange} vertical line of the prediction phase in kWh. The upper x-axis displays the equivalent CO\textsubscript{2} emissions in grams (gCO\textsubscript{2}e), calculated based on the 2023 world average.}
    \Description[Test]{Test}
    \label{fig:stages}
\end{figure}

The energy consumption during the training stage of a recommender system model is, on average, ten times higher than predicting the top-n items for all users (0.41 vs. 0.04 kWh, \cref{fig:stages}).

Energy use varies across different stages of a recommender system.
Typically, training a model consumes more energy than generating predictions. 
For instance, deep learning models like \emph{DGCF} and \emph{MacridVAE} consume 73 and 44 times more energy during training than the prediction phase. 
On the other hand, traditional models such as \emph{SVD} and \emph{NMF} consume less energy during training compared to generating predictions, 70 and 19 times less, respectively.

Although the training phase consumes, in general, ten times more energy, our findings suggest that the high energy cost of training can, in some cases, be offset by energy savings during prediction.
For instance, while \emph{MacridVAE} from \emph{RecBole} uses nearly 4,400 times more energy during training than the \emph{SVD} implementation from \emph{RecPack} (1.75 kWh vs. 0.0004 kWh), it consumes seven times less energy during prediction (0.04 kWh vs. 0.28 kWh). 
This means that after just seven prediction cycles, \emph{SVD} would exceed the energy consumption of training \emph{MacridVAE} despite the significant difference in training energy usage.

\subsubsection{The Energy Consumption of an Experimental Pipeline}\label{ssec:pipeline}
Based on our experiments, we estimate the energy consumption of a representative 2023 recommender systems experimental pipeline to be 151.2 kWh.

Through the paper analysis described in \cref{paper_review}, we found that a 2023 recommender systems experimental pipeline includes, on average, seven recommender systems models. 
Additionally, the models' performance is, on average, evaluated on three datasets. 
Furthermore, a representative experimental pipeline performs hyperparameter optimization through grid search on 16 configurations per model. 
Since one model consumes, on average\footnoteref{note1}, 0.45 kWh (\cref{fig:avgpowcons}, left), the energy consumption of an experimental pipeline is calculated as follows:  
 \[ 7 \times 3 \times 16 \times 0.45 \, kWh = 151.2 \, kWh.\]

\subsubsection{The Energy Consumption of a Paper}
\label{interview}
Based on experiments and our paper study (\cref{paper_review}), we estimate the energy consumption of a representative 2023 paper to be 6,048 kWh.

We define a representative 2023 recommender systems paper as one that follows contemporary methodological standards, as observed in our paper study (\cref{paper_review}), and employs the representative experimental pipeline we characterized (\cref{rep_pipeline}).
The energy consumption estimation of 151.2 kWh per recommender systems experimental pipeline only accounts for the direct energy consumption during the experimental run (\cref{ssec:pipeline}).
The estimation excludes energy costs for preliminary activities such as model prototyping, initial test runs, data collection, data preprocessing, debugging, and potential re-running of experiments due to pipeline errors.
Therefore, to approximate the total energy impact of a recommender systems paper, we account for these additional energy costs by introducing an additional factor. 

We interviewed the authors of Elliot \cite{DBLP:conf/sigir/AnelliBFMMPDN21}, RecPack \cite{10.1145/3523227.3551472}, LensKit \cite{ekstrand2020lenskit}, and recommender systems practitioners \cite{10.1145/3604915.3609498} asking them to estimate a factor of the energy consumption overhead of a recommender systems paper compared to running the experimental pipeline once.
Their median answer was 40.
We multiplied the energy consumption of experimental pipelines accordingly: 
\[ 151.2 \, kWh \times 40 = 6,048 \, kWh \]

Across the three repeated runs, we observed high measurement consistency, with energy consumption varying by an average of 3.5\%\footnote{Differences in average consumption may vary due to external factors such as temperature, device characteristics, and other uncontrolled conditions.} due to uncontrollable factors, e.g., thermal fluctuations. 
Hyperparameter configurations showed a smaller impact, with 2.2\%\footnote{This value is an average of our tested hyperparameter configurations. It may vary depending on hardware, hyperparameter ranges, algorithmic implementation, and random initialization.} average variation in energy use between configurations. These results confirm both the reliability of our smart-plug method and that model architecture choices dominate energy efficiency over hyperparameter tuning.

\subsection{Energy Consumption and Performance Trade-Off}
\label{performance-tradeoff}

\begin{figure}
    \centering
    \resizebox{0.65\linewidth}{!}{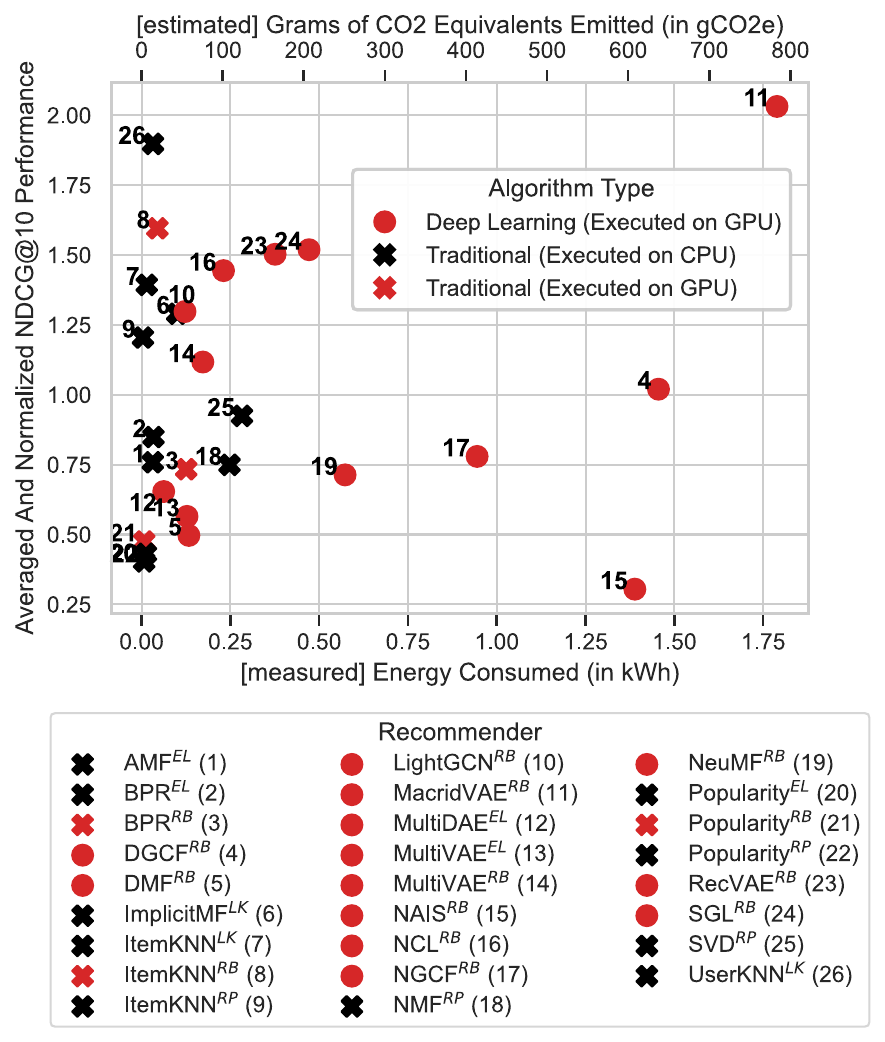}
    \caption{
    Total energy consumption (in kWh) vs. averaged and normalized \emph{nDCG@10} performance. Data points in \textbf{\emph{black}} represent models running on CPUs and \textbf{\emph{red}} for running on GPUs. A cross represents a traditional, and a dot is a deep learning model. The \emph{nDCG@10} is normalized within each dataset to ensure uniform impact and averaged across all twelve included datasets. The upper x-axis shows the gCO\textsubscript{2}e emissions, calculated using the 2023 world average.}
    \Description[Test]{Test}
    \label{fig:relation}
\end{figure}

Our results demonstrate that higher energy consumption does not necessarily lead to better performance when using default hyperparameters (\cref{fig:relation}).
Two of the three top-performing models (\emph{UserKNN$^{LK}$} and \emph{ItemKNN$^{RB}$}) are traditional nearest neighbor models that both consume on average around 0.040 kWh for a single run on one dataset (average of twelve datasets used for 2023 experiments). 
In contrast, the third, a deep learning model (\emph{MacridVAE$^{RB}$}) consumes, on average, 45 times more energy at 1.8 kWh per run on one dataset.

In general, our results show that deep learning models consume eight times more energy than traditional models for a single run, on average, on one dataset (0.09 kWh vs. 0.68 kWh) without achieving a higher average and normalized \emph{nDCG@10} score.
One comparison that further illustrates the energy consumption and performance trade-off between deep learning and traditional models is the comparison between the \emph{MacridVAE$^{RB}$} (deep learning) and the \emph{UserKNN$^{LK}$} (traditional) models. 
While both models achieve around the same normalized and averaged \emph{nDCG@10} performance (1.8 vs. 1.7), \emph{MacridVAE$^{RB}$} consumes almost 60 times more energy (1.79 kWh vs. 0.03 kWh).

As widely known and highlighted by current research, hyperparameter optimization and randomness impact the model performance \cite{10.1145/3604915.3609488, WegmethVPB23}.
We acknowledge that tuning hyperparameters and repeating experiments could alter our performance results. 
However, hyperparameter optimization will increase the energy consumption disparity between deep learning and traditional models. 
For instance, if we optimize the hyperparameters of \emph{MacridVAE$^{RB}$} and \emph{UserKNN$^{LK}$} through a grid search with 16 configurations and repeat the process with five different seeds, \emph{MacridVAE$^{RB}$} would consume 143.2 kWh, while \emph{UserKNN$^{LK}$} would only use 2.4 kWh (factor of 60).

The energy consumption difference between deep learning and traditional models is not solely due to GPU usage.
While GPUs consume more energy than CPUs, implementation and complexity also play a role. 
For instance, \emph{ItemKNN$^{RB}$} on a GPU does not consume significantly more energy than CPU-based \emph{KNN} counterparts (\cref{fig:relation}).

\subsection{Carbon Footprint and Trends}\label{results_carbon}

\subsubsection{Carbon Footprint Analysis of ACM RecSys 2023 Experiments}\label{energy RecSys23}

Based on our experiments, we estimate that running all ACM RecSys 2023 full paper experiments emitted 782.5 metric tonnes of CO\textsubscript{2} equivalents. Our CO\textsubscript{2}e estimation is based on the number of submissions and the conversion factors from kWh to gCO\textsubscript{2}e.

The carbon footprint of all ACM RecSys 2023 full papers is closely tied to the number of submissions received. 
The conference accepted 47 full papers out of 269 submissions. 
Since the submissions involved running experiments, every submission added to the total carbon footprint of the ACM RecSys 2023 experiments. 
Consequently, we account for all 269 submissions in our analysis\footnote{We acknowledge that the specific methodological details and precise reasons for rejection for each submission are not available. 
However, we assert that papers submitted to the ACM RecSys conference generally adhere to a shared experimental paradigm, often utilizing common datasets and model architectures, independent of the decision to accept or reject the paper.}.

Our carbon footprint estimation is further based on the \emph{world average} conversion factor of 481 gCO\textsubscript{2}e per kWh \cite{ember2024carbon}.
We estimate that a full paper experimental pipeline consumes, on average, 6,048 kWh (\cref{RQ1}).
With the conversion factor of 481 gCO\textsubscript{2}e per kWh, the carbon emissions of the 2023 ACM RecSys conference experiments in metric tonnes of CO\textsubscript{2}e are calculated as follows:
\[ 6,048 kWh \times 481 gCO\textsubscript{2}e \, per \,
kWh = 2,909 \, kgCO\textsubscript{2}e \, (per \, paper) \]
\[ 2,909 \, kgCO\textsubscript{2}e \, (per \, paper) \, \times 269 \, (submissions) = 782.5 \, TCO\textsubscript{2}e \]
To illustrate, 782.5 TCO\textsubscript{2}e is the equivalent of 338 passenger flights from New York (USA) to Melbourne (Australia) \cite{https://doi.org/10.1002/advs.202100707}. 
Or the amount of CO\textsubscript{2}e that, on average, one tree sequesters in 71,100 years \cite{UKGov2020}.

\subsubsection{Geographic Impact on Carbon Footprint}\label{location_impact}
Variations in energy generation across different geographical locations affect carbon footprint by as much as 1200\% (45 vs. 535 \gcotwoe). 

Each location has its own kWh to \gcotwoe conversion factor, which reflects the carbon intensity of its energy generation.
The conversion factor is based on the energy sources of the respective geographical location. 
For instance, we ran our experiments in Sweden, a geographical location mainly utilizing renewable energy sources such as wind and hydropower \cite{ZHONG2021812}. 
Unlike this geographical location, some regions depend on coal or other fossil fuels \cite{pressburger2022comprehensive}. 
As a result, the conversion factor from kWh to gCO\textsubscript{2} can be twelve times lower than, for example, that of Asia (45 vs. 535) \cite{ember2024carbon}.

If all experiments from ACM RecSys 2023 had been conducted in Sweden like ours, the carbon emission estimation would have been reduced to 74 TCO\textsubscript{2}e (90\% less than 782.5 TCO\textsubscript{2}e).
In contrast, running the experiments in Asia, our estimation would have increased our carbon emission estimation by 93.1 TCO\textsubscript{2}e (782.5 vs. 875.6 TCO\textsubscript{2}e)
Since no paper from ACM RecSys 2023 reported the data center or location where the experiments were conducted, we used the \emph{world average} conversion factor of 481 gCO\textsubscript{2}e per kWh \cite{ember2024carbon} to convert kWh to gCO\textsubscript{2}e and estimate the carbon emissions.

The carbon emissions from experiments are not solely determined by the location.
Some data centers operate mainly on renewable energy regardless of their location. 
For instance, \emph{Amazon} reports that their data centers used for \emph{AWS} cloud services predominantly utilize renewable sources\footnote{\url{https://sustainability.aboutamazon.com/products-services/the-cloud}}. 
Consequently, these specific data centers may have a conversion factor from kWh to gCO\textsubscript{2}e that differs from the general rate of their location.

On the other hand, networking costs also play a role. For example, if researchers in Asia use data centers in Norway, they must account for the energy used for data traffic, though this is much harder to quantify.

\subsubsection{How Hardware Choices Impact Carbon Footprint}\label{hardware_carbon}
\begin{figure}
    \centering
    \resizebox{0.75\linewidth}{!}{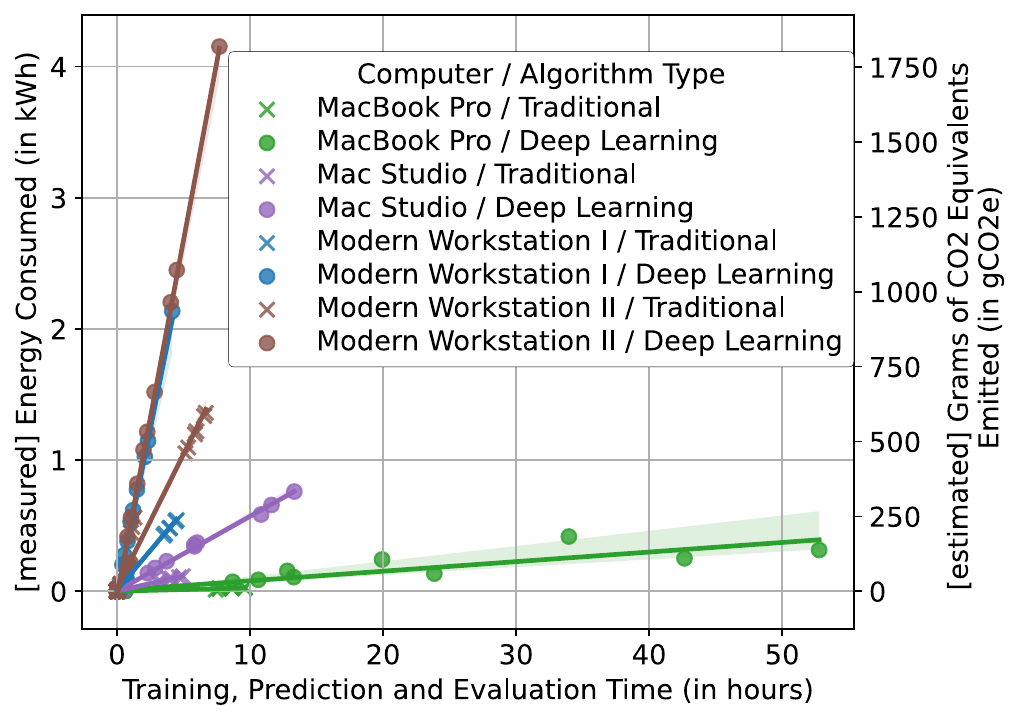}
    \caption{The relationship between energy consumption (in kWh) and runtime, including training, prediction, and evaluation phases (in seconds). Each data point represents a model applied to one of the twelve datasets. The linear functions illustrate the linear regression models for the respective groups of data points. The right-hand y-axis displays the corresponding gCO\textsubscript{2}e emissions, calculated using the 2023 world average.}
    \Description[Test]{Test}
    \label{figure3}
\end{figure}

\begin{table}
\caption{Power draw of hardware used to compare efficiency.}
\begin{tabular}{l|r|r}
\toprule
Computer and Year & Idle Power Draw & Max Power Draw \\
\midrule
Modern Workstation I; 2023 & 116 Watt ($\sigma$= 2.03 Watt) & 547 Watt\\
Modern Workstation II; 2023 & 180 Watt ($\sigma$= 4.35 Watt) & 580 Watt \\
Mac Studio; 2022 & 18 Watt ($\sigma$= 0.71 Watt) & 73 Watt\\
MacBook Pro; 2020 & 1 Watt ($\sigma$= 0.02 Watt) & 21 Watt\\
Legacy Workstation; 2013 & 67 Watt ($\sigma$= 3.30 Watt) & 139 Watt\\
\bottomrule
\end{tabular}
\label{tab:energy}
\end{table}

The carbon emissions of recommender systems experiments are influenced by more factors than geographical location, energy sources, model types, and dataset characteristics. 
The type of hardware executing experiments can also affect the carbon emissions by a factor of up to ten. 
For example, the same experiments emit 14.4 gCO\textsubscript{2}e when executed on an M1 MacBook Pro but 163.5 gCO\textsubscript{2}e when executed on Modern Workstation I with an NVIDIA GPU.

Various hardware components, architectures, and cooling methods affect the energy consumption of recommender system experiments.  
Based on our experiments, \emph{Modern Workstation I} with an NVIDIA GPU consumes, on average, five times more energy compared to M1 Ultra Mac Studio (0.33 vs. 0.07 kWh) and ten times more energy compared to M1 MacBook Pro (0.33 vs. 0.03 kWh). 
More precisely, under full load, \emph{Modern Workstation II} draws approximately 28 times more power than the M1 MacBook Pro, drawing 580 W compared to just 21 W (\cref{tab:energy}). 
Even at idle, the power draw of Modern Workstation II remains significantly higher, drawing 116 times more power than the M1 MacBook Pro, at 116 W versus 1 W (\cref{tab:energy}).

Different hardware types not only affect the energy consumption but also the running time of a recommender systems experiment. 
For instance, while \emph{Modern Workstation I} and \emph{Modern Workstation II} use, on average, ten times more energy than M1 MacBook Pro, both \emph{Modern Workstation I} and \emph{Modern Workstation II} complete the experiments, on average, in only one-third of the time (\cref{figure3}). 
Therefore, saving energy by running an experiment on an M1 MacBook Pro is possible with an increase in running time.

On the other hand, similar hardware types exhibit comparable energy consumption and runtime. 
Both \emph{Modern Workstation I} and \emph{Modern Workstation II} are equipped with an NVIDIA RTX 3090 GPU, leading to similar energy consumption (0.33 vs. 0.34 kWh) and runtime (0.57 vs. 0.53 hours) for deep learning models.
However, despite sharing the same GPU, the two devices differ in other key components, such as CPUs, operating systems, cooling solutions, and memory capacity. 
These variations result in small performance and energy consumption differences, particularly for traditional models, which we observe by measuring all hardware components.

Overall, our results show a linear trend between energy consumption and runtime across various hardware types (\cref{figure3}). 
Although the slopes of the graphs vary, a consistent linear pattern is evident. 
This relationship suggests that a longer runtime is associated with increased energy consumption.

Even though both \emph{Modern Workstation I} and \emph{Modern Workstation II} consume more energy, we run experiments on them because not all models are compatible with Apple's ARM architecture.

\subsubsection{Comparing RecSys Carbon Footprints: 2013 vs. 2023}
\begin{figure}
    \centering
    \resizebox{0.75\linewidth}{!}{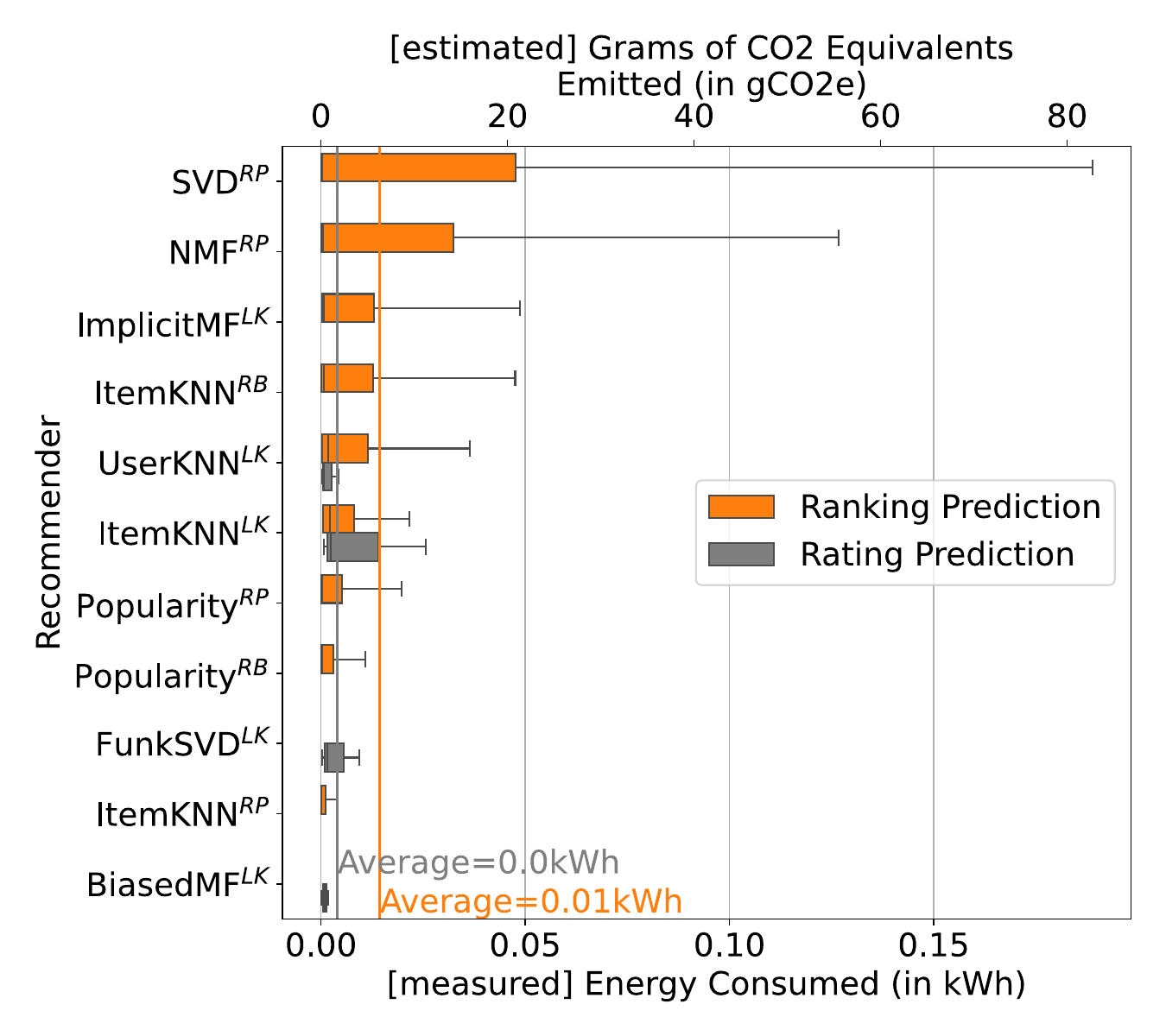}
    \caption{Average energy consumption of traditional models executed on 2013 hardware across seven datasets. The \emph{orange} vertical line indicates the average energy consumption in kWh for ranking predictions and the \emph{orange} for rating prediction tasks. The upper x-axis shows the gCO\textsubscript{2}e emissions, calculated using the 2023 world average. Not every model is suited for rating and ranking prediction tasks; therefore, not every model displays two boxplots.}
    \Description[Test]{Test}
    \label{figure4}
\end{figure}

\begin{figure}
    \centering
    \resizebox{0.75\linewidth}{!}{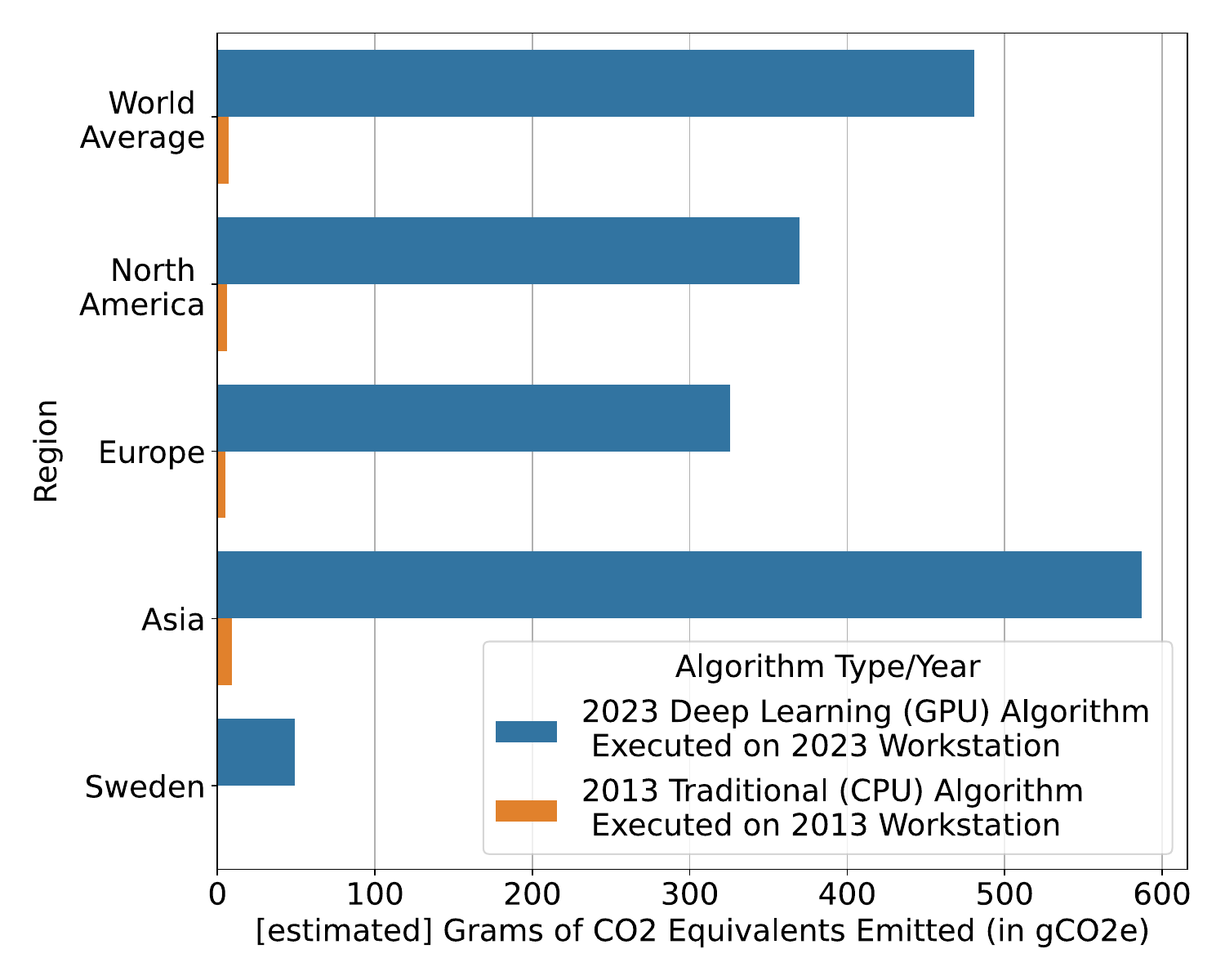}
    \caption[Test]{The regional variations in gCO\textsubscript{2}e per recommender system model type. The x-axis displays the gCO\textsubscript{2}e, while the y-axis categorizes by region. Blue bars represent the emissions from a representative deep learning model executed on the Modern Workstation I hardware, and orange bars represent those from a traditional model run on 2013 hardware. The gCO\textsubscript{2}e are calculated using the respective annual conversion factors, reflecting changes in gCO\textsubscript{2} per kWh over the decade.}
    \Description[Test2]{}
    \label{figure5}
\end{figure}

The carbon footprint of recommender systems experiments in 2023 compared to 2013 has, on average, increased by a factor of 42 (7.09 gCO\textsubscript{2}e in 2013 vs. 294,9 gCO\textsubscript{2}e in 2023, \cref{figure5}, World Average). 

Simple ranking prediction models used in 2013 recommender system experiments consumed, on average, one-fifth as much energy per run on one 2013 dataset using 2013 hardware (0.1 kWh, see \cref{figure4}) compared to similar models in 2023 executed on 2023 hardware (0.51 kWh, see \cref{fig:avgpowcons}).
In 2013, no model used, on average, more than 0.2 kWh per run on one dataset using 2013 hardware (\cref{figure4}). 
In contrast, all deep learning models in 2023 consume, on average, more than 0.2 kWh per run on one dataset with 2023 hardware. 
The most energy-intensive models in 2023 use more than six kWh per run (\cref{fig:avgpowcons}).
While rating prediction was frequently used in 2013 recommender system experiments, the energy consumption for rating prediction is, due to prediction times, lower than for ranking prediction tasks (\cref{figure4}).

The shift towards more complex deep learning models such as \emph{DGCF} or \emph{MacridVAE} and larger datasets like \emph{Yelp-2018}, compared to simpler models like \emph{Popularity} or \emph{ItemKNN} and smaller datasets such as \emph{Hetrec-LastFM} used in 2013, can, in the worst case, increase energy consumption by more than 100,000 times.
For example, running the \emph{Popularity} model on \emph{Hetrec-LastFM} and hardware from 2013 requires only 0.000049 kWh while running \emph{DGCF} on dataset \emph{Yelp-2018} required 6.6 kWh on 2023 hardware, i.e., a factor of around 134,000. 

The increased usage of clean energy sources and improved hardware efficiency in 2023, compared to 2013, does not compensate for the increase in carbon emissions due to the use of deep learning models and larger datasets. 
A deep learning model running on a 2023 dataset using 2023 hardware emits, on average, 42 times more CO\textsubscript{2}e than a traditional model on a 2013 dataset using 2013 hardware (7.09 gCO\textsubscript{2}e in 2013 vs. 294,9 gCO\textsubscript{2}e in 2023, \cref{figure5}, World Average). 
This estimation already includes the benefit of a better kWh to gCO\textsubscript{2}e conversion factors, improved by 62 gCO\textsubscript{2}e due to the more frequent use of clean energy resources \cite{ember2024carbon}.

We highlight that our analysis used hardware and models from 2013, but not the exact software implementations used in the original papers. 
It was infeasible to retrieve old, unmaintained, non-centrally hosted software.
Consequently, many experiments likely used custom implementations of models. 
In comparison, standardized software libraries could potentially change the efficiency, suggesting that the observed disparity in energy consumption between traditional and deep learning models might diverge.

%% file: figure1.pdf_tex
\begingroup%
  \makeatletter%
  \providecommand\color[2][]{%
    \errmessage{(Inkscape) Color is used for the text in Inkscape, but the package 'color.sty' is not loaded}%
    \renewcommand\color[2][]{}%
  }%
  \providecommand\transparent[1]{%
    \errmessage{(Inkscape) Transparency is used (non-zero) for the text in Inkscape, but the package 'transparent.sty' is not loaded}%
    \renewcommand\transparent[1]{}%
  }%
  \providecommand\rotatebox[2]{#2}%
  \newcommand*\fsize{\dimexpr\f@size pt\relax}%
  \newcommand*\lineheight[1]{\fontsize{\fsize}{#1\fsize}\selectfont}%
  \ifx\svgwidth\undefined%
    \setlength{\unitlength}{648bp}%
    \ifx\svgscale\undefined%
      \relax%
    \else%
      \setlength{\unitlength}{\unitlength * \real{\svgscale}}%
    \fi%
  \else%
    \setlength{\unitlength}{\svgwidth}%
  \fi%
  \global\let\svgwidth\undefined%
  \global\let\svgscale\undefined%
  \makeatother%
  \begin{picture}(1,1.11111111)%
    \lineheight{1}%
    \setlength\tabcolsep{0pt}%
    \put(0,0){\includegraphics[width=\unitlength,page=1]{figure1.pdf}}%
  \end{picture}%
\endgroup%

%% file: figure1_2.pdf_tex
\begingroup%
  \makeatletter%
  \providecommand\color[2][]{%
    \errmessage{(Inkscape) Color is used for the text in Inkscape, but the package 'color.sty' is not loaded}%
    \renewcommand\color[2][]{}%
  }%
  \providecommand\transparent[1]{%
    \errmessage{(Inkscape) Transparency is used (non-zero) for the text in Inkscape, but the package 'transparent.sty' is not loaded}%
    \renewcommand\transparent[1]{}%
  }%
  \providecommand\rotatebox[2]{#2}%
  \newcommand*\fsize{\dimexpr\f@size pt\relax}%
  \newcommand*\lineheight[1]{\fontsize{\fsize}{#1\fsize}\selectfont}%
  \ifx\svgwidth\undefined%
    \setlength{\unitlength}{720bp}%
    \ifx\svgscale\undefined%
      \relax%
    \else%
      \setlength{\unitlength}{\unitlength * \real{\svgscale}}%
    \fi%
  \else%
    \setlength{\unitlength}{\svgwidth}%
  \fi%
  \global\let\svgwidth\undefined%
  \global\let\svgscale\undefined%
  \makeatother%
  \begin{picture}(1,1)%
    \lineheight{1}%
    \setlength\tabcolsep{0pt}%
    \put(0,0){\includegraphics[width=\unitlength,page=1]{figure1_2.pdf}}%
  \end{picture}%
\endgroup%

%% file: figure8.pdf_tex
\begingroup%
  \makeatletter%
  \providecommand\color[2][]{%
    \errmessage{(Inkscape) Color is used for the text in Inkscape, but the package 'color.sty' is not loaded}%
    \renewcommand\color[2][]{}%
  }%
  \providecommand\transparent[1]{%
    \errmessage{(Inkscape) Transparency is used (non-zero) for the text in Inkscape, but the package 'transparent.sty' is not loaded}%
    \renewcommand\transparent[1]{}%
  }%
  \providecommand\rotatebox[2]{#2}%
  \newcommand*\fsize{\dimexpr\f@size pt\relax}%
  \newcommand*\lineheight[1]{\fontsize{\fsize}{#1\fsize}\selectfont}%
  \ifx\svgwidth\undefined%
    \setlength{\unitlength}{432bp}%
    \ifx\svgscale\undefined%
      \relax%
    \else%
      \setlength{\unitlength}{\unitlength * \real{\svgscale}}%
    \fi%
  \else%
    \setlength{\unitlength}{\svgwidth}%
  \fi%
  \global\let\svgwidth\undefined%
  \global\let\svgscale\undefined%
  \makeatother%
  \begin{picture}(1,1)%
    \lineheight{1}%
    \setlength\tabcolsep{0pt}%
    \put(0,0){\includegraphics[width=\unitlength,page=1]{figure8.pdf}}%
  \end{picture}%
\endgroup%

%% file: figure3.pdf_tex
\begingroup%
  \makeatletter%
  \providecommand\color[2][]{%
    \errmessage{(Inkscape) Color is used for the text in Inkscape, but the package 'color.sty' is not loaded}%
    \renewcommand\color[2][]{}%
  }%
  \providecommand\transparent[1]{%
    \errmessage{(Inkscape) Transparency is used (non-zero) for the text in Inkscape, but the package 'transparent.sty' is not loaded}%
    \renewcommand\transparent[1]{}%
  }%
  \providecommand\rotatebox[2]{#2}%
  \newcommand*\fsize{\dimexpr\f@size pt\relax}%
  \newcommand*\lineheight[1]{\fontsize{\fsize}{#1\fsize}\selectfont}%
  \ifx\svgwidth\undefined%
    \setlength{\unitlength}{424.49639893bp}%
    \ifx\svgscale\undefined%
      \relax%
    \else%
      \setlength{\unitlength}{\unitlength * \real{\svgscale}}%
    \fi%
  \else%
    \setlength{\unitlength}{\svgwidth}%
  \fi%
  \global\let\svgwidth\undefined%
  \global\let\svgscale\undefined%
  \makeatother%
  \begin{picture}(1,1.1778356)%
    \lineheight{1}%
    \setlength\tabcolsep{0pt}%
    \put(0,0){\includegraphics[width=\unitlength,page=1]{figure3.pdf}}%
  \end{picture}%
\endgroup%

%% file: figure2.pdf_tex
\begingroup%
  \makeatletter%
  \providecommand\color[2][]{%
    \errmessage{(Inkscape) Color is used for the text in Inkscape, but the package 'color.sty' is not loaded}%
    \renewcommand\color[2][]{}%
  }%
  \providecommand\transparent[1]{%
    \errmessage{(Inkscape) Transparency is used (non-zero) for the text in Inkscape, but the package 'transparent.sty' is not loaded}%
    \renewcommand\transparent[1]{}%
  }%
  \providecommand\rotatebox[2]{#2}%
  \newcommand*\fsize{\dimexpr\f@size pt\relax}%
  \newcommand*\lineheight[1]{\fontsize{\fsize}{#1\fsize}\selectfont}%
  \ifx\svgwidth\undefined%
    \setlength{\unitlength}{489.17202759bp}%
    \ifx\svgscale\undefined%
      \relax%
    \else%
      \setlength{\unitlength}{\unitlength * \real{\svgscale}}%
    \fi%
  \else%
    \setlength{\unitlength}{\svgwidth}%
  \fi%
  \global\let\svgwidth\undefined%
  \global\let\svgscale\undefined%
  \makeatother%
  \begin{picture}(1,0.70635576)%
    \lineheight{1}%
    \setlength\tabcolsep{0pt}%
    \put(0,0){\includegraphics[width=\unitlength,page=1]{figure2.pdf}}%
  \end{picture}%
\endgroup%

%% file: figure4.pdf_tex
\begingroup%
  \makeatletter%
  \providecommand\color[2][]{%
    \errmessage{(Inkscape) Color is used for the text in Inkscape, but the package 'color.sty' is not loaded}%
    \renewcommand\color[2][]{}%
  }%
  \providecommand\transparent[1]{%
    \errmessage{(Inkscape) Transparency is used (non-zero) for the text in Inkscape, but the package 'transparent.sty' is not loaded}%
    \renewcommand\transparent[1]{}%
  }%
  \providecommand\rotatebox[2]{#2}%
  \newcommand*\fsize{\dimexpr\f@size pt\relax}%
  \newcommand*\lineheight[1]{\fontsize{\fsize}{#1\fsize}\selectfont}%
  \ifx\svgwidth\undefined%
    \setlength{\unitlength}{648bp}%
    \ifx\svgscale\undefined%
      \relax%
    \else%
      \setlength{\unitlength}{\unitlength * \real{\svgscale}}%
    \fi%
  \else%
    \setlength{\unitlength}{\svgwidth}%
  \fi%
  \global\let\svgwidth\undefined%
  \global\let\svgscale\undefined%
  \makeatother%
  \begin{picture}(1,0.88888889)%
    \lineheight{1}%
    \setlength\tabcolsep{0pt}%
    \put(0,0){\includegraphics[width=\unitlength,page=1]{figure4.pdf}}%
  \end{picture}%
\endgroup%

%% file: figure5.pdf_tex
\begingroup%
  \makeatletter%
  \providecommand\color[2][]{%
    \errmessage{(Inkscape) Color is used for the text in Inkscape, but the package 'color.sty' is not loaded}%
    \renewcommand\color[2][]{}%
  }%
  \providecommand\transparent[1]{%
    \errmessage{(Inkscape) Transparency is used (non-zero) for the text in Inkscape, but the package 'transparent.sty' is not loaded}%
    \renewcommand\transparent[1]{}%
  }%
  \providecommand\rotatebox[2]{#2}%
  \newcommand*\fsize{\dimexpr\f@size pt\relax}%
  \newcommand*\lineheight[1]{\fontsize{\fsize}{#1\fsize}\selectfont}%
  \ifx\svgwidth\undefined%
    \setlength{\unitlength}{720bp}%
    \ifx\svgscale\undefined%
      \relax%
    \else%
      \setlength{\unitlength}{\unitlength * \real{\svgscale}}%
    \fi%
  \else%
    \setlength{\unitlength}{\svgwidth}%
  \fi%
  \global\let\svgwidth\undefined%
  \global\let\svgscale\undefined%
  \makeatother%
  \begin{picture}(1,0.8)%
    \lineheight{1}%
    \setlength\tabcolsep{0pt}%
    \put(0,0){\includegraphics[width=\unitlength,page=1]{figure5.pdf}}%
  \end{picture}%
\endgroup%

%% file: content/discussion.tex
\section{Discussion}\label{discussion}
Our analysis reveals the substantial environmental impact of full papers from the 2023 ACM Conference on Recommender Systems. 
We found that training deep learning models consumes eight times more energy for a single run than traditional machine learning approaches, without delivering superior performance. 
Moreover, the carbon footprint of recommender systems experiments has increased by a factor of 42 over the past decade. 
In this section, we discuss the key factors contributing to this environmental impact, as identified in our study. 
We also explore emerging sustainability trends in recommender systems for 2024, particularly the growing use of transformer-based models such as large language models (LLMs), which further escalate computational demands and energy consumption.

\subsection{Carbon Footprint Impact Factors}
In this section, we discuss the three largest single carbon footprint impact factors based on our results: energy production methods, hardware efficiency, and wasteful computation.
This should not imply that these are the only impact factors, but we believe that their significance makes these a good starting point for carbon footprint minimization efforts.
Furthermore, we emphasize that the energy consumption of recommender systems is just one of several contributors to their overall carbon footprint.
The life cycle of computing resources for running recommender systems and other resource-intensive tasks comes with other contributors such as e-waste \cite{PERKINS2014286}, natural resource extraction \cite{Ensmenger2018TheEH}, heat generation in data centers \cite{Park2022CurrentAP}, deforestation for data center construction \cite{SOVACOOL2022102493}, and water consumption \cite{li2023makingaithirstyuncovering}.
The share of environmental impact of each factor is different for most researchers and challenging, if not impossible, to accurately measure.
However, this shows that the carbon footprint of recommender systems experiments is higher than we portray with our calculations, as we consider no other measurements but the carbon footprint from energy generation.
This is significant since our conservative estimate of 2,909 kilograms of CO\textsubscript{2} equivalents per research paper, ignoring impact factors other than energy consumption, is already alarmingly high.

\subsubsection{Energy Production Methods}
We demonstrate how the location of computing resources affects the carbon footprint of recommender system experiments (\cref{location_impact}).
The amount of emitted CO\textsubscript{2} equivalents at each location depends on the energy mix used in that location.
However, this is a macro view providing an average value.
For example, while Sweden no longer runs coal power plants, it still runs a small portion of its grid with energy produced through coal combustion \cite{iea_sweden_coal}. 
Therefore, a researcher could unknowingly use energy originating from fossil fuels, even in Sweden, if they ignore the energy mix used by their local energy or data center provider.
Conversely, while, for example, Germany powers a considerable part of its grid with energy produced from coal combustion, renewable energy is widely available for personal and commercial use \cite{iea_germany_energy_mix}.
As a result, it is, in many cases, possible to choose an energy source that produces fewer CO\textsubscript{2} equivalents, no matter the location. This can be accomplished by either changing the energy provider as a private or commercial user, or by selecting a data center that uses green energy sources.

To illustrate the difference in emissions of different energy production methods, we show an excerpt of the so-called life cycle emissions of different energy production technologies provided by the Intergovernmental Panel on Climate Change (IPCC) \cite{schlomer2014_annexIII} in Table \ref{ghg}.
The life cycle emissions contain all emissions of the given production method, e.g., direct emissions, infrastructure emissions, and others.
The highest median emissions are from coal, while the lowest are from nuclear, geothermal, and wind.
The difference in terms of the median between these production methods is a factor of roughly 70, e.g., running the same recommender systems experiments with energy produced from coal combustion emits 70 times more CO\textsubscript{2} equivalents than using nuclear, geothermal, or wind energy instead.

\begin{table}[htbp]
\centering
\caption{Life cycle emissions of energy production technologies in gCO\textsubscript{2}e/kWh, sorted by median value descending \cite{schlomer2014_annexIII}.}
\begin{tabular}{l|l}
\toprule
\textbf{Options} & \textbf{Lifecycle Emissions (Min/Median/Max)} \\ 
\midrule
Coal—PC & 740/820/910 \\ 
Biomass—cofiring & 620/740/890 \\ 
Gas—Combined Cycle & 410/490/650 \\ 
Biomass—dedicated & 310/330/420 \\ 
Solar PV—rooftop & 26/41/60 \\ 
Solar PV—utility & 18/38/180 \\ 
Hydropower & 1.0/24/220 \\
Concentrated Solar Power & 8.8/22/63 \\ 
Nuclear & 3.7/12/110 \\
Geothermal & 6/11/90 \\ 
Wind onshore & 7.0/11/56 \\ 
Wind offshore & 8.0/11/25 \\ 
\bottomrule
\end{tabular}
\label{ghg}
\end{table}

\subsubsection{Hardware Efficiency}
Our results show that specific hardware, e.g., the Mac M1 chip, can reduce energy consumption by 90\% compared to consumer-grade GPUs, e.g., NVIDIA RTX 3090 (\cref{hardware_carbon}).
Although the M1 chip is slower than other hardware we compared against, most experiments can be run in parallel, offsetting the time loss at the cost of a more extensive infrastructure.
Furthermore, modern hardware becomes more energy-efficient and time-efficient with each generation.
Modern hardware optimized for high performance per watt ensures that fewer resources are wasted.
For example, the NVIDIA Tesla A100 SXM2\footnote{\url{https://www.nvidia.com/en-us/data-center/a100/}} 
is capable of almost 2.5 times as many TFLOPS as the NVIDIA Tesla V100 SXM\footnote{\url{https://www.nvidia.com/en-gb/data-center/tesla-v100/}} but consumes only 30\% more energy.

The caveats to hardware efficiency are that hardware production also incurs a carbon footprint and that upgrading recommender systems infrastructure may incur heavy financial costs.
Therefore, especially when upgrading legacy infrastructure, the carbon footprint reduction is not immediate.
Instead, the impact of the carbon footprint of hardware production is amortized over time with hardware usage.
According to Apple\footnote{\url{https://www.apple.com/environment/pdf/products/desktops/Mac_Studio_PER_March2025.pdf}}, the life cycle carbon footprint of the latest 2025 Mac Studio is 276 kgCO\textsubscript{2}e, down from 382 kgCO\textsubscript{2}e in the previous generation. 
Taking into account only production and transportation, a 2025 Mac Studio generates 163 kgCO\textsubscript{2}e.
Hence, the production of 20 Mac Studio generates as much CO\textsubscript{2}e as a 2023 recommender systems research paper, according to our experiments.
Upgrading legacy hardware on-premises can be circumvented by choosing a data center provider that is equipped with efficient hardware.
However, this may lead to the data center requiring more hardware and, therefore, incurring the aforementioned carbon footprint from hardware production.

\subsubsection{Wasteful Computation}
The outcome of the interview with our peers (\cref{interview}), as well as reflecting on our own experience writing research papers that require computation for experiments, tells us that we often repeat experiments unnecessarily.
Sometimes, repeating experiments is necessary, i.e., to confirm a result or eliminate the impact of randomness.
Other times, experiments are repeated because there was a bug in the code that produced faulty results, or the code forcefully stopped due to an unhandled exception.
In any case, failed experiments consume energy and, therefore, generate a carbon footprint.
Many researchers and practitioners may not be aware of how many times they repeat experiments due to bugs and thereby waste computation.
For example, they may not care to repeat failed experiments because they consider it cheaper in terms of time investment than implementing the necessary checks for these errors.

To illustrate the potential impact of wasteful computation, consider that we estimate the energy consumption of a representative 2023 recommender systems experimental pipeline executed once with complete success (\cref{ssec:pipeline}).
This estimate excludes data preprocessing efforts such as cleaning and feature engineering, exploratory runs like hyperparameter sweeps and ablation studies, failed experiments including crashed training jobs and diverged models, redundant executions caused by bugs or hardware failures, idle resource time from GPU or CPU under-utilization, debugging overhead such as profiling and sanity checks, and other hidden inefficiencies like storage bottlenecks and network latency.
After prompting our interviewees to consider this overhead, the median answer was a multiplier of 40 (\cref{interview}), with responses as low as 5 and as high as 300.
We can see that this overhead varies significantly depending on the individual and the project.
Nevertheless, our findings show that the overhead caused by wasteful computations is substantial.

\subsection{Trends in 2024 and Beyond}
We consider papers from 2013 and 2023 in our paper study (\cref{paper_review}) to build the representative pipelines used in our experiments.
To extend on these insights and provide an outlook for the future, we performed a paper study for the year 2024 with the same parameters.
In \cref{tab:comparison_24}, we present the trends from papers accepted at ACM RecSys 2024 compared to 2023 to discuss how recommender systems experimental design and reporting changes in one year and what this means for the sustainability of recommender systems.
We make some notable observations: (1) More papers disclose the used hardware and the majority switched to the next generation of GPUs, (2) the used libraries and experimental pipelines, including hyperparameter optimization, dataset type and quantity, and prediction task, are unchanged, and (3) significantly more papers share reproducibility packages.

\begin{table}[h!]
\centering
\caption{Comparison of recommender systems research papers from ACM RecSys 2023 and 2024.}
\label{tab:comparison_24}
\begin{tabular}{l|l|l}
\toprule
\textbf{Statistic}                    & \textbf{ACM RecSys 2023}                   & \textbf{ACM RecSys 2024}                     \\ \midrule
Number of accepted papers                      & 47                              & 58                                \\
Papers detailing hardware             & 15 papers (32\%)      & 25 papers (43\%)          \\
Most common hardware & NVIDIA V100 GPU & NVIDIA A100 GPU \\
Most common libraries                 & PyTorch, TensorFlow, RecBole   & PyTorch, TensorFlow, RecBole      \\
Data splitting technique              & Holdout (20 papers, 42\%)       & Holdout (17 papers, 29\%)         \\
Hyperparameter optimization           & 34 papers (72\%)              & 36 papers (62\%)                  \\
Most common optimization technique    & Grid search (31 papers, 66\%)   & Grid search (27 papers, 47\%)     \\
Most common evaluation metric                    & nDCG (32 papers, 68\%)    & nDCG (36 papers, 62\%)            \\
Most common prediction task                      & Top-n ranking prediction (94\%)        & Top-n ranking prediction (91\%)   \\
Papers sharing code                   & 29 papers (62\%)                  & 44 papers (76\%)                  \\
Mean datasets used in each paper               & 2.85                            & 2.81                              \\
Most popular datasets                 & Amazon, MovieLens, LastFM      & Amazon, MovieLens, Yelp  \\
Includes transformer models & 21 papers (45\%) & 32 papers (55\%) \\
\bottomrule
\end{tabular}
\end{table}

We built the representative experimental pipeline based on our paper study (\cref{paper_review}) and included deep learning recommender systems models, excluding transformer models\footnote{We use transformer models as an umbrella term for model architectures using transformers, including LLMs.}.
In 2013, transformer models did not exist.
Recall that Word2Vec was only introduced in 2013 \cite{mikolov2013efficientestimationwordrepresentations}, and modern transformer models were introduced in 2017 \cite{vaswani2023attentionneed}.
By 2023, transformer models were already widely accessible, and conferences, e.g., ACM RecSys 2023, accepted research papers exploring their development and application for recommendation tasks.
We found that, at ACM RecSys 2023, 21 (45\%) of accepted full papers contain transformer models in the methodology.
One year later, in 2024, 32 (55\%) papers, i.e., more than half of the papers accepted, are about transformer models in recommender systems.
There already exist comprehensive peer-reviewed surveys, e.g., at ACM TOIS, specifically about LLMs in recommender systems \cite{10.1145/3678004,li-etal-2024-large,wu2024surveylargelanguagemodels}, emphasizing the significance of this trend and the recommender systems community's acceptance of this new algorithmic direction.
Researchers further show that transformer models lead to higher computational costs than other deep-learning approaches and, therefore, a higher carbon footprint \cite{DBLP:journals/corr/abs-2104-10350,strubell2019energy}.
Therefore, we find that, given the current trends in recommender systems experimental design, our carbon footprint estimate for a recommender systems paper is a lower bound for today's research landscape.
Furthermore, our results confirm that computational cost, e.g., time, affects the carbon footprint proportionally.
Consequently, given the increasing popularity of transformer models in recommender systems research, we are confident that the carbon footprint will significantly increase in the next few years due to model size and complexity, possibly more than in the decade from 2013 to 2023.

However, we do not advocate abandoning deep learning, including transformer models, but aim to raise awareness about the environmental impact of the trend toward deep learning-focused research. 
We encourage researchers and practitioners to document the experimental pipelines, computational overheads, hardware, and energy consumption in their publications. 
These details can help understand the environmental impact, highlight the energy demands, and reveal potential areas for energy efficiency improvements and reproducibility of recommender systems experiments.

%% file: content/guidelines.tex
\section{Practical Guidelines for Green Recommender Systems}

\subsection{Practical Guidelines for Planning and Executing Green Recommender Systems}\label{guidelines_planning}
This section provides guidelines to assist researchers and practitioners, collectively referred to as developers, in planning and executing green recommender systems. 
The guidelines focus on minimizing recommender systems environmental impact using energy-efficient hardware, green energy sources, measuring and reporting energy consumption, and carbon offset programs.
Reducing the environmental impact of recommender systems experiments starts with avoiding unnecessary computations. 

Unnecessary computation should be avoided to lower energy consumption and decrease carbon emissions (\cref{energy_to_carbon}).
This can be achieved with energy-aware recommender systems pipeline design, notably by (1) employing test-driven development, e.g., with pytest\footnote{\url{https://docs.pytest.org/en/stable/}}, to prevent errors that result in re-runs, (2) choosing models and datasets that align with the research goals, (3) prototyping in Python or Rust but running production code with efficient, low-level languages, e.g., C, and (4) using rigorously tested and optimized libraries \cite{10.1145/3591109} like RecBole \cite{recbole[1.2.0]}, RecPack \cite{10.1145/3523227.3551472}, and LensKit \cite{ekstrand2020lenskit} instead of implementing algorithms from scratch, avoiding redundant implementation efforts, streamlining validation, and minimizing the risk of errors.

Developers should prioritize energy-efficient hardware, e.g., the Apple M1 chip or its successors, to reduce energy consumption by up to 90\% compared to the NVIDIA 3090 GPU (\cref{hardware_carbon}).
Although energy-efficient hardware may require more time to compute, most experiments can be run in parallel, offsetting the time loss at the cost of a more extensive infrastructure.
Furthermore, modern hardware becomes more energy-efficient and time-efficient with each generation, e.g., the NVIDIA Tesla A100 SXM2\footnote{\url{https://www.nvidia.com/en-us/data-center/a100/}} 
completes 2.5 times as many TFLOPS as the NVIDIA Tesla V100 SXM\footnote{\url{https://www.nvidia.com/en-gb/data-center/tesla-v100/}} at 30\% increased energy consumption.

Green energy sources, e.g., renewable or nuclear, should be prioritized over fossil fuels, as the difference in carbon emissions can be up to 70 times (\cref{ghg}).
To run computations with local resources, developers should pick an energy provider with a green energy mix.
When selecting a provider for cloud computing, developers should consider the data center's commitment to sustainable infrastructure, as green data centers use renewable energy, efficient cooling, and carbon offset programs. 
Developers should review the environmental credentials and verify whether providers fulfill sustainability promises. 

The energy consumption of experiments should be measured and reported in a standard unit, e.g., kilowatt-hours.
With local computational resources, developers should measure energy consumption using tools such as, e.g., EMERS \cite{10.1007/978-3-031-87654-7_8} or CodeCarbon \cite{benoit_courty_2024_11171501}.
In cloud computing, some providers, e.g., AWS\footnote{\url{https://aws.amazon.com/aws-cost-management/aws-customer-carbon-footprint-tool/}}, offer energy consumption measurements.
The energy consumption of each executed dataset-model combination should be transparent, especially in research that presents novel algorithms. 
This ensures that the recommender systems community avoids favoring energy-intensive solutions that come with significant environmental costs despite offering minimal performance improvements (\cref{performance-tradeoff}) \cite{10.1145/3604915.3608840}.
Furthermore, the report should be as granular and comprehensive as possible, separating the energy consumption by different phases of experimentation, e.g., data processing, training, and prediction.
Finally, the report should cover the energy consumption for all experiments, including those not part of the final manuscript.

Energy-efficient models such as ItemKNN, BPR, and LightGCN, or pre-trained models, should be prioritized to reduce energy consumption while maintaining high performance.
For example, energy-efficient models often deliver effective results for collaborative filtering tasks without the energy costs of complex deep learning models (\cref{performance-tradeoff}) \cite{10.1145/3604915.3608840}.
Alternatively, leveraging pre-trained models minimizes energy consumption by eliminating the need for resource-intensive training, allowing fine-tuning with smaller datasets instead.

Carbon offset programs help mitigate the environmental impact of unavoidable emissions, whether from experiments, legacy infrastructure, or travel.
While awareness and well-planned pipelines help to reduce the carbon footprint, green recommender systems will have a carbon footprint.
Carbon offsetting involves supporting projects that remove or prevent the equivalent amount of CO\textsubscript{2} from entering the atmosphere, such as reforestation or renewable energy initiatives. 
For example, donating to tree-planting programs counterbalances emissions from recommender system experiments. 
However, carbon offsetting via tree-planting programs may have drawbacks: carbon sequestration occurs slowly over a tree’s lifetime, global reforestation potential is limited, and poorly executed projects may inadvertently damage existing ecosystems \cite{tree_planting}.
It is crucial to emphasize that carbon offsets serve as a complementary measure rather than a primary solution. 
Their most appropriate use is in addressing emissions that cannot yet be eliminated through direct reduction strategies. 

\subsection{Practical Guidelines for Documenting Green Recommender Systems Research}\label{guidelines_documenting}
In this section, we propose a framework designed to assist authors in systematically documenting the environmental impact of their research. 
This framework emphasizes reporting detailed information on computing resources, energy usage, carbon emissions, and strategies implemented to reduce resource consumption while aligning with global sustainability goals. 
To support its practical application, we provide a checklist to help authors incorporate all relevant documentation into their research where appropriate in the Appendix (\cref{checklist}). 
Adherence to this framework ensures the work is transparent and accountable regarding its environmental impact.

\subsubsection{Hardware Documentation}
Authors should provide detailed descriptions of the hardware used in their research. 
This includes comprehensive specifications of hardware components, such as the type and model of CPUs, GPUs, TPUs, memory capacity, storage devices, e.g., SSDs or HDDs, networking equipment, and any other relevant hardware used in the project.
For instance, specifying that the research utilized an AMD Ryzen Threadripper CPU, an NVIDIA Tesla V100 GPU, and NVMe SSD storage offers insights into the energy efficiency of individual components. 
Additionally, authors should indicate whether their hardware was purchased for their research or if existing resources were utilized, pointing out that the production and transportation of new hardware increases the environmental footprint.
This underscores the ecological implications of hardware production and emphasizes the value of minimizing environmental impact by reusing existing resources.

\subsubsection{Energy Mix \& Consumption Assessment}
Authors should report the total energy consumed in kilowatt-hours for all computational tasks, quantifying the overall environmental impact of their research.
For example, stating that their entire research project consumed 150 kWh highlights its resource intensity. 
Furthermore, they should describe the energy mix used to power their computations, including the proportion of renewable and non-renewable sources, to provide an insight into the environmental impact of the energy used. 
If authors use different computational resources, e.g., cloud-based systems and on-premises servers, they should specify different energy mixes separately.
For example, they could specify that their on-premise servers' energy mix is 80\% solar and 20\% coal, but their cloud-based systems' energy mix is 100\% hydropower.
Finally, authors should specify whether they actively selected renewable energy sources, such as using a cloud provider certified for 100\% renewable energy, demonstrating a commitment to sustainability. 

\subsubsection{Planning \& Executing Experiments}
Authors should describe strategies implemented to reduce resource consumption, such as algorithmic optimizations or efficient hyperparameter searches, demonstrating efforts to minimize environmental impact.
For example, authors could detail how early stopping criteria during training significantly reduced energy consumption. 
Furthermore, authors should assess whether the granularity of their energy and resource measurements is detailed enough to provide meaningful insights into the various stages of the experiments, such as data processing, training, and inference, specifying whether measurements were made at the level of individual tasks or only reported in aggregate. 
For instance, measuring energy usage separately for data pre-processing, training, and prediction tasks allows a better understanding of where the most significant energy consumption occurs and enables targeted optimizations.
Authors should also report the time required for significant computational tasks, such as model training or simulations, to provide context for understanding resource usage and workflow efficiency.
Finally, they should disclose whether and how many additional experiments or exploratory runs were performed but not included in the final results, ensuring a more accurate understanding of total resource consumption.

\subsubsection{Energy-Efficiency Considerations}
Authors should assess the energy cost-effectiveness of their approach by comparing energy consumption to performance gains, such as calculating energy-per-unit improvement in accuracy, to determine whether the benefits justify the energy usage. 
Additionally, authors should clarify whether users can balance performance and energy consumption, enabling energy savings while still achieving acceptable results for specific use cases. 
This could involve reducing model complexity, lowering computational precision, or utilizing less powerful hardware. 
Furthermore, authors should discuss how energy consumption scales with increases in input data, model size, or workload, evaluating the environmental feasibility of their approach for larger-scale applications.
Finally, they should evaluate whether their experiments can be reproduced using a scaled-down computational setup without significant loss of validity, promoting accessibility and sustainability.

\subsubsection{Carbon Emission Assessment}
Authors should calculate and report the carbon emissions of their research by considering energy consumption, the energy mix, and other factors such as emissions from travel, hardware production, and related activities, providing a comprehensive assessment of the environmental impact. 
Additionally, they should indicate whether steps were taken to offset this footprint, such as purchasing carbon credits or supporting renewable energy projects, demonstrating accountability and proactive sustainability measures. 
Furthermore, authors should identify sustainability goals aligned with their research, e.g., the UNFCCC (\cref{introduction}) and the UN SGDs (\cref{unsdg}), and explain how their work contributes to achieving these goals, for instance, by optimizing models to reduce energy consumption or promoting environmentally conscious computing practices.

%% file: content/conclusions.tex
\section{Conclusions}
We reveal that the energy consumption of an average recommender systems research paper is approximately 6,048 kWh (\textbf{RQ1}).
Deep learning models consume, on average, eight times more energy than traditional models without achieving higher performance with default hyperparameters (\textbf{RQ2}). 
The carbon footprint of recommender systems experiments has increased significantly, with experiments from 2023 emitting approximately 42 times more CO\textsubscript{2}e when compared to experimental pipelines from 2013 (\textbf{RQ3}).

The environmental impact of recommender systems can be minimized through efficient hardware, careful model and dataset selection, and sustainable energy sources.
Energy-efficient hardware, such as Apple's M1 chip, can cut energy consumption by up to 90\% compared to dedicated GPUs like the NVIDIA 3090 (\cref{hardware_carbon}). 
Similarly, model and dataset choices significantly influence runtime and energy use, with variations as high as a factor of 1,444 (e.g., DGCF on Hetrec-LastFM vs. DGCF on Yelp-2018, \cref{consumption_per_dataset}). 
The energy mix plays a critical role, as regions like Sweden, which rely on an energy mix dominated by renewable energy sources, exhibit up to 12 times less carbon emissions compared to regions where the energy mix is dependent on fossil fuel-based sources (\cref{location_impact}). 

The recommender systems community has yet to explore the development of green recommender systems --- recommender systems explicitly designed to minimize their energy consumption and carbon footprint.
This gap persists despite the growing urgency to address global sustainability challenges.
Additionally, the carbon footprint of recommender systems has increased significantly over the past decade and will likely keep increasing over the following years.
This increase is driven by the shift toward more complex deep learning models and the expansion of dataset sizes over the last ten years. 
Recent trends involving LLMs further exacerbate the issue.

To assist researchers and practitioners in minimizing the carbon footprint of recommender systems, we propose practical guidelines for planning, implementing, and documenting green recommender systems (\cref{guidelines_planning} and \cref{guidelines_documenting}).
Awareness and transparency are essential for identifying potential areas for energy efficiency improvements and catalyzing the development of more sustainable practices.
The guidelines are based on our results and focus on environmentally conscious decision-making and transparency about the environmental impact of recommender systems.
By following our guidelines, researchers can reduce the carbon cost of recommender systems research while still advancing the state of the art in recommendation technologies.

To assess and mitigate the environmental impact of our research, we quantified the carbon emissions from our experiments and took steps to offset the emissions produced.
We conducted the experiments on Modern Workstation I, Mac Studio, and Legacy Workstation in Gothenburg\footnote{\url{https://www.svk.se/om-kraftsystemet/om-elmarknaden/elomraden/}}, Sweden, which consumed approximately 6,100 kWh of energy,
The carbon intensity of energy generation was 35 gCO\textsubscript{2}e/kWh, and 74\% of the energy came from renewable sources, primarily wind and hydropower.
Furthermore, we conducted experiments on Modern Workstation II and MacBook Pro in Siegen, Germany, which consumed approximately 500 kWh of energy.
The carbon intensity of energy generation was 466 gCO\textsubscript{2}e/kWh, and 48\% of the energy came from renewable sources, primarily solar and wind, while the non-renewable fraction came from coal.
Our total energy consumption corresponds to 446.5 kgCO\textsubscript{2}e in carbon emissions. 
We planted 42 trees with One Tree Planted to offset our carbon emissions\footnote{According to One Tree Planted, one tree absorbs 10kg of CO\textsubscript{2} per year. Therefore, we offset our carbon emissions over 11 months.}.

Our analysis provides foundational and actionable insights into the environmental impact of the recommender systems community. As such, the conclusions drawn in this work are grounded in the context of recommender systems experiments. We posit that some of the patterns and relationships uncovered here may be applicable in other areas of machine learning, and we encourage researchers in those fields to conduct similar analyses.

\subsection{Future Work}

While our study quantifies the energy consumption of recommender systems, several promising research directions emerge that could be addressed in future work.\newline 
\indent\textit{Model Diversity:} Our work primarily examines collaborative filtering, but future studies should investigate other approaches, including content-based filtering, knowledge-aware models, context-aware recommenders, reinforcement learning-based systems, and hybrid architectures. 
The increased complexity of these approaches may lead to increased energy consumption compared to collaborative filtering. 
Transformer-based models, in particular, demand further analysis due to their high computational costs and growing adoption.\newline 
\indent\textit{Evaluation Metrics:} We assess models using ranking-based accuracy metrics. 
Future work should incorporate additional dimensions such as diversity, fairness, and novelty, while exploring their trade-offs with energy efficiency. 
Developing optimization strategies that balance these metrics with sustainability remains an open challenge.\newline 
\indent\textit{Hyperparameter Optimization:} Our analysis reveals a 2.2\% average variation in energy consumption across model configurations, demonstrating the sensitivity of energy efficiency to hyperparameter choices. While we intentionally used default configurations to assess fundamental energy-performance trade-offs and minimize unnecessary computational overhead, this finding suggests that systematic hyperparameter optimization could yield deeper insights into efficiency trade-offs. Future work should investigate optimal configurations that balance computational demands with model accuracy.\newline 
\indent\textit{Stakeholder Perspectives:} While we focus on single-stakeholder systems, multi-stakeholder and group recommenders present unique challenges that warrant investigation from an energy-efficiency standpoint, particularly in scenarios with conflicting objectives.\newline 
\indent\textit{Modality Scope:} Finally, we analyze interaction data only, but multi-modal recommenders, integrating text, images, or audio, are increasingly prevalent. 
Their additional complexity likely increases energy demands, making them a critical area for future sustainability research.\newline 
\indent\textit{Adoption Incentives:} While our paper provides guidelines for green recommender systems, implementing them often incurs overhead without immediate benefits, other than reducing environmental impact, which some may not find rewarding enough.
Therefore, it may require additional incentives to convince recommender systems researchers and practitioners to adopt green recommender systems guidelines.
Future work should explore incentive mechanisms, such as certification systems, carbon-aware benchmarking, or regulatory frameworks, to encourage academia and industry adoption of sustainable practices.\newline 
\indent\textit{Transferring Knowledge:} Our work provides an isolated view on the energy consumption of recommender systems and does not consider the application of energy reduction techniques from other applied machine learning fields. We find that there is a lot to learn from sustainability research in other fields. Future research should address the application of energy consumption and carbon footprint reduction techniques from other fields in recommender systems.

This study serves as a call to action for researchers and practitioners to embrace the principles of green recommender systems and work towards a more sustainable future with AI-powered personalization.

%% file: content/appendix.tex
\appendix
\section{Appendix}

\subsection{Aligning Recommender Systems with United Nations Sustainable Development Goals}\label{unsdg}
Recommender systems should be aligned with the United Nations Sustainable Development Goals (SDGs), focused on sustainability, social responsibility, and accountable use of technology.
Here, we discuss approaches that support goals like SDG 12---Responsible Consumption and Production, and SDG 13---Climate Action, demonstrating how technological advancements in recommender systems can be steered towards sustainability.

Modern recommender systems often use deep learning models and train complex neural networks on large datasets to deliver highly personalized recommendations, but they come with significant environmental costs.
On a global average, one person produced 6.6 tCO\textsubscript{2}e per year in 2023 \cite{crippa_ghg_2024}.
This is only twice our estimated figure of 3.3 tCO\textsubscript{2}e per recommender systems research paper in 2023.
Conservatively assuming that the recommender systems community writes 10,000 new papers\footnote{A Google Scholar search for recommender systems articles (search term: ``recommender system''|``recommender systems''|``recommendation system''|``recommendation systems'') published in 2023 returns 16,200 results.} containing experiments in a year, running experiments to write recommender systems research papers produces as much CO\textsubscript{2}e as almost 5,000 people in one year.
However, while the CO\textsubscript{2}e emissions per year per capita are relatively stable over time, we project the emissions of recommender systems experiments to increase.
This environmental cost prompts questions about whether high-emission models are necessary for every use case.
We show that ``Good old-fashioned AI'' models, like neighborhood-based methods, can offer competitive recommendation accuracy without deep learning’s energy demands.
For some applications, the incremental accuracy of deep learning may not justify its carbon footprint. 
For instance, recommending digital content, e.g., music or movies, has a low environmental impact even if recommendations are suboptimal, i.e., skipping to the next track or selecting a different movie causes negligible CO\textsubscript{2} emissions. 
However, in fashion e-commerce, poor recommendations can lead to high environmental costs from shipping and returns. 
Accurate recommendations could reduce unnecessary shipments, directly supporting SDG 12 by encouraging responsible consumption.
To mitigate environmental impact, practitioners can explore energy-efficient models, optimize resources, and consider carbon offsets. 
Additionally, recommender systems can align with SDG 13 by prioritizing sustainable options, such as local products or eco-friendly items, fostering environmentally conscious behaviors across different application contexts.

\subsection{Checklist for Documenting Green Recommender Systems Research}\label{checklist}
We discuss practical guidelines for documenting green recommender systems research in \cref{guidelines_documenting}.
Here, we provide a checklist with key questions to guide authors in documenting key environmental impact factors of their research.
These questions encourage the inclusion of detailed information about computing resources, energy use, and carbon emissions, as well as the strategies employed to minimize resource consumption and align with global sustainability goals, e.g., the UNFCCC (\cref{introduction}) and the UN SGDs (\cref{unsdg}).
We suggest that authors of research papers do not explicitly state these questions in their papers, nor add them to their appendix in a checklist.
Instead, authors should include the answers in their text where appropriate.
Answering the questions ensures that research papers are transparent about environmental impact factors.
Additionally, conference organizers could adopt and propagate this checklist to encourage a stronger focus on green research practices across submissions.
We provide a numbered list of the questions, followed by a guide on how to answer the questions.

\begin{enumerate}
    \item Which computing environment and location were used?
\item Was new hardware acquired for the project?
\item Which hardware is used?
\item What is the energy mix?
\item Were renewable energy sources explicitly chosen for computations?
\item How will the energy consumption of the presented approach scale with size?
\item Is the presented approach energy cost-effective?
\item Can users compromise energy consumption for performance?
\item What is the duration of computational tasks?
\item Were additional experiments conducted beyond what is reported?
\item What is the total energy consumption of the project?
\item What is the carbon footprint of the paper?
\item Which strategies to minimize resource consumption were used?
\item Was the carbon footprint of the work offset?
\item Is the granularity of the measurements sufficient?
\item Are the reported results reproducible with fewer resources?
\item Does the work align with sustainability goals?
\item What is the work's expected societal or practical impact?
\item How does this work contribute to advancing sustainable research practices?
\end{enumerate}

\subsubsection{Which computing environment and location were used?}
Authors should describe the computing resources and environment used for their work, such as cloud-based systems, on-premises servers, or personal devices, along with the location of these resources.
This explanation should include the rationale for their choice, considering cost, scalability, convenience, and sustainability, as well as the geographic region, which can impact energy efficiency due to cooling requirements and the local energy mix.  
For example, selecting a cloud provider located in a region with abundant renewable energy and using GPUs optimized for energy efficiency can significantly reduce the environmental impact of the research.

\subsubsection{Was new hardware acquired for the project?}
Authors should specify whether new hardware was purchased or acquired for the project.  
This information highlights the environmental impact of hardware production and whether the research relied on pre-existing resources.
For example, purchasing a new GPU for the study increases the resource footprint, whereas utilizing an existing system demonstrates resource efficiency.

\subsubsection{Which hardware is used?}
Authors should provide detailed information about the hardware components employed in their research, including the make and model of CPUs, GPUs, TPUs, or other specialized accelerators.
In addition to processors, this should cover memory capacity, storage devices, e.g., SSDs or HDDs, networking equipment, and any other relevant hardware used in the computational process.
For instance, specifying a combination of a high-performance CPU, e.g., AMD Ryzen Threadripper, a powerful GPU, e.g., NVIDIA Tesla V100, and fast storage, e.g., NVMe SSD, helps clarify the infrastructure's capacity, energy consumption, and performance characteristics.  

\subsubsection{What is the energy mix?}
The energy mix powering the computations should be described, including the proportion of renewable and non-renewable sources. 
This provides insight into the environmental impact of the energy used. 
For instance, authors could specify that 80\% of the energy came from solar and wind while the remaining 20\% came from coal or natural gas.  

\subsubsection{Were renewable energy sources explicitly chosen for computations?}
Authors should specify whether they actively selected renewable energy sources to power their computations. 
This includes detailing any data center energy policies or renewable energy certifications. 
For example, using a cloud provider certified for 100\% renewable energy demonstrates a commitment to sustainability. 

\subsubsection{How will the energy consumption of the presented approach scale with size?}
Authors should discuss how energy consumption is expected to grow as the size of the input data, model, or workload increases. 
This helps evaluate whether the approach is environmentally feasible for larger-scale applications. 
For instance, an approach whose energy usage grows sub-linearly with data size may be more sustainable than one with exponential growth. 

\subsubsection{Is the presented approach energy cost-effective?}
Authors should assess the energy cost-effectiveness of their approach by comparing energy consumption to performance gains. 
This analysis highlights whether the achieved improvements justify the energy usage. 
For example, calculating energy-per-unit improvement in accuracy provides a precise measure of cost-effectiveness. 

\subsubsection{Can users compromise energy consumption for performance?}
Authors should clarify whether users of their novel approach can sacrifice performance for reduced energy consumption and, if so, how this trade-off can be made.  
This could include options such as reducing the model's complexity, lowering the precision of computations, or using less powerful hardware, which may lead to lower energy use at the cost of performance metrics like accuracy or speed.  
For example, authors could explain how a user might run the model with fewer layers or lower precision arithmetic to decrease energy consumption while maintaining acceptable results for specific use cases. 

\subsubsection{What is the duration of computational tasks?}
Authors should report the time required for significant computational tasks, such as model training or simulations. 
This provides context for understanding resource usage and workflow efficiency. 
For instance, noting that a model took 48 hours to train offers a baseline for estimating energy consumption. 

\subsubsection{Were additional experiments conducted beyond what is reported?}
Authors should disclose whether additional experiments or exploratory runs were performed but not included in the final results. 
This ensures a more accurate understanding of the total resource consumption. 
For example, reporting that 20 exploratory experiments were conducted before selecting the final model improves transparency. 

\subsubsection{What is the total energy consumption of the project?}
Authors should report the total energy consumed in kilowatt-hours (kWh) for all computational tasks. 
This quantifies the overall environmental impact of the research. 
For example, stating that the project consumed 150 kWh highlights its resource intensity. 

\subsubsection{What is the carbon footprint of the paper?}
Authors should calculate and report the carbon footprint of their research, using the measured energy consumption and the energy mix to estimate emissions associated with computational tasks.
In addition to energy usage, authors should consider other relevant factors, such as emissions from travel, hardware production, and any other activities that contribute to the environmental impact of the research.
For example, the carbon footprint calculation could include travel to conferences or the energy consumed during hardware manufacturing, providing a more comprehensive view of the paper’s environmental impact.

\subsubsection{Which strategies to minimize resource consumption were used?}
Authors should describe strategies implemented to reduce resource consumption, such as algorithmic optimizations or efficient hyperparameter searches.
This shows efforts to minimize the environmental impact of the research.
For example, authors could detail how early stopping criteria during training significantly reduced energy use.

\subsubsection{Was the carbon footprint of the work offset?}
Authors should indicate whether steps were taken to offset the carbon footprint of their research, such as purchasing carbon credits.
This demonstrates accountability for environmental impact.
For instance, offsetting 100 kgCO2e by contributing to a wind energy project, purchasing carbon credits, or paying for a service that plants trees reflects a proactive sustainability measure.

\subsubsection{Is the granularity of the measurements sufficient?}
Authors should assess whether the granularity of their energy and resource measurements is detailed enough to provide meaningful insights into the various stages of the experiments, such as data processing, training, and inference.
This includes specifying whether measurements were made at the level of individual tasks, such as energy consumed during model training versus inference, or if only aggregate measurements were reported.
For example, measuring energy usage separately for training, data pre-processing, and inference tasks allows a better understanding of where the most significant energy consumption occurs and enables targeted optimizations.

\subsubsection{Are the reported results reproducible with fewer resources?}
Authors should evaluate whether their experiments can be reproduced using a scaled-down computational setup without significant loss of validity. 
This promotes accessibility and sustainability in research. 
For example, showing that a smaller model achieves similar results supports resource-efficient practices. 

\subsubsection{Does the work align with sustainability goals?}
Authors should identify sustainability goals that align with their research and explain how their work contributes to achieving these goals.  
This could include goals related to energy efficiency, reducing environmental impact, or promoting responsible resource usage, among others (\cref{unsdg}).  
For example, authors could describe how their research on optimizing recommender systems contributes to reducing energy consumption or how their work on algorithmic efficiency supports responsible computing practices.  

\subsubsection{What is the work's expected societal or practical impact?}
Authors should justify the environmental costs of their research by outlining its potential societal or practical benefits. 
This ensures the value of the work outweighs its environmental impact. 
For instance, a model that improves healthcare diagnostics while consuming modest energy has significant practical benefits. 

\subsubsection{How does this work contribute to advancing sustainable research practices?}
Authors should reflect on how their work contributes to sustainable research practices. 
This may include proposing methodologies or frameworks that inspire environmentally conscious research. 
For example, introducing a more energy-efficient training pipeline sets a precedent for future studies. 

\subsection{Energy Consumption Tables}\label{ect}
\begin{center}
\begin{table}[!ht]
\caption{Statistics of energy consumption across datasets. Values represent mean, median, minimum, and maximum energy consumption (in kWh) of 17 models on 12 datasets.}
\begin{tabular}{l|r|r|r|r}
\toprule
 Dataset & \makecell[tl]{Average Energy \\ Consumption} & \makecell[tl]{Median Energy \\ Consumption} & \makecell[tl]{Minimum Energy \\Consumption} & \makecell[tl]{Maximum Energy \\ Consumption} \\
\midrule
AMZ-Books & 0.704400 & 0.327300 & 0.012800 & 2.665200 \\
AMZ-CDs-And-Vinyl & 0.362500 & 0.193300 & 0.008400 & 1.865600 \\
AMZ-Electronics  & 0.813100 & 0.275400 & 0.007800 & 6.474400 \\
AMZ-Sports-And-Outdoors & 0.605400 & 0.478500 & 0.008700 & 2.897300 \\
AMZ-Toys-And-Games & 0.775400 & 0.315300 & 0.009500 & 3.320300 \\
Gowalla & 0.682800 & 0.252300 & 0.009000 & 3.063900 \\
Hetrec-LastFM & 0.001200 & 0.000600 & 0.000000 & 0.006900 \\
MovieLens-100K  & 0.001800 & 0.000600 & 0.000000 & 0.010300 \\
MovieLens-1M & 0.063000 & 0.004900 & 0.000800 & 0.613500 \\
MovieLens-Latest-Small  & 0.002400 & 0.000500 & 0.000000 & 0.017900 \\
Retailrocket & 0.031300 & 0.016200 & 0.002200 & 0.123400 \\
Yelp-2018 & 1.364900 & 0.594300 & 0.030200 & 6.586100 \\
\bottomrule
\end{tabular}
\end{table}
\end{center}

\begin{center}
\begin{table}[!ht]
\caption{Statistics of energy consumption across recommenders. Values represent mean, median, minimum, and maximum energy consumption (in kWh) of 17 models executed on 12 datasets.}
\begin{tabular}{l|r|r|r|r}
\toprule
Recommender & \makecell[tl]{Average Energy \\ Consumption} & \makecell[tl]{Median Energy \\ Consumption} & \makecell[tl]{Minimum Energy \\Consumption} & \makecell[tl]{Maximum Energy \\ Consumption} \\
\midrule
AMF$^{EL}$ & 0.031500 & 0.036300 & 0.000300 & 0.099800 \\
BPR$^{RB}$ & 0.125600 & 0.081600 & 0.000300 & 0.559900 \\
DGCF$^{RB}$ & 1.455300 & 1.291300 & 0.004600 & 6.586100 \\
DMF$^{RB}$ & 0.133200 & 0.099200 & 0.001200 & 0.718600 \\
ImplicitMF$^{LK}$ & 0.100400 & 0.120400 & 0.000300 & 0.261400 \\
ItemKNN$^{RB}$ & 0.044500 & 0.055800 & 0.000100 & 0.093700 \\
LightGCN$^{RB}$ & 0.122300 & 0.120800 & 0.000800 & 0.495400 \\
MacridVAE$^{RB}$ & 1.788900 & 2.265400 & 0.000600 & 3.979100 \\
MultiDAE$^{EL}$ & 0.062300 & 0.048300 & 0.000100 & 0.195100 \\
MultiVAE$^{RB}$ & 0.172500 & 0.157800 & 0.000100 & 0.530800 \\
NAIS$^{RB}$ & 1.388700 & 1.164700 & 0.006900 & 6.474400 \\
NCL$^{RB}$ & 0.230700 & 0.225500 & 0.001500 & 0.887200 \\
NGCF$^{RB}$ & 0.944700 & 0.265200 & 0.001400 & 5.279500 \\
NMF$^{RP}$ & 0.247600 & 0.236200 & 0.000200 & 0.549300 \\
NeuMF$^{RB}$ & 0.572900 & 0.575600 & 0.000700 & 2.135500 \\
Popularity$^{RB}$ & 0.007500 & 0.008100 & 0.000000 & 0.030200 \\
RecVAE$^{RB}$ & 0.375900 & 0.351900 & 0.000500 & 1.146300 \\
SGL$^{RB}$ & 0.471500 & 0.458100 & 0.001900 & 1.929800 \\
SVD$^{RP}$ & 0.282700 & 0.320000 & 0.000100 & 0.687000 \\
\bottomrule
\end{tabular}
\end{table}
\end{center}